\documentclass{iopjournal}
\usepackage{natbib}[numbers]
\usepackage{amsmath,amsthm,amssymb,bm}
\usepackage{parskip}
\newcommand{\abs}[1]{\left| #1 \right|}
\newcommand{\floor}[1]{\left\lfloor #1 \right\rfloor}
\newcommand{\norm}[1]{\left\lVert #1 \right\rVert}
\newcommand{\df}{\mathrm{d}}
\newcommand{\dd}[2]{\frac{\mathrm{d} #1}{\mathrm{d} #2}}

\theoremstyle{definition}

\theoremstyle{remark}

\theoremstyle{plain}

\theoremstyle{plain}

\theoremstyle{plain}

\theoremstyle{plain}

\theoremstyle{plain}

\newcommand{\rad}{v} 
\newcommand{\Wbar}{\mathsf{W}} 
\newcommand{\Bobar}{\bm{\mathsf{B}}_0} 
\newcommand{\Bbar}{\bm{\mathsf{B}}} 

\begin{document}

\articletype{Paper} 

\title{Statistical equilibrium model for stellarators}

\author{M. Ruth$^{1,3}$, J. W. Burby$^{1,2,*}$, W. Sengupta$^{4}$, A. Brown$^{4,5}$}

\affil{$^1$Department of Physics, University of Texas at Austin, Austin, TX  78712 USA}

\affil{$^2$Institute for Fusion Studies, University of Texas at Austin, Austin, TX  78712 USA}

\affil{$^3$Oden Institute, University of Texas at Austin, Austin, TX  78712 USA}

\affil{$^4$Department of Astrophysical Sciences, Princeton University, Princeton, NJ 08543 USA}

\affil{$^5$Princeton Plasma Physics Laboratory, Princeton, NJ 08540 USA}

\affil{$^*$Author to whom any correspondence should be addressed.}

\email{joshua.burby@austin.utexas.edu}

\keywords{equilibrium}

\begin{abstract}
In three dimensional toroidal domains without symmetry, the standard magnetohydrodynamic (MHD) equilibrium model used for magnetic confinement fusion does not generally support smooth solutions. 
Instead, solutions have singular plasma currents on resonant magnetic surfaces that violate the MHD assumption of length-scale separation, further leading to the non- or slow convergence of numerical approximations under refinement.
In this work, we present an improved equilibrium principle derived from a statistical model for plasma fluctuations.
Instead of being static, we assume that the plasma magnetic field is ergodically and rapidly fluctuating relative to the MHD time scale.
By averaging the resulting force, we derive a variational equilibrium problem for the statistical mean magnetic field which depends on fluctuation variance.
Then, through asymptotics, numerical simulations, and a Grad-Shafranov type argument, we show that the variational principle supports smooth solutions for specific fluctuation statistics chosen to minimally modify the standard equilibrium modeling paradigm.
Physically, this model smooths singular current sheets with a length scale determined by the magnetic field fluctuations.
\end{abstract}

\section{Introduction\label{introduction}}
One of the promising fusion energy concepts is the stellarator \cite{imbert-gerard_introduction_2024}.
Stellarators work by using strong non-axisymmetric toroidal magnetic fields to control the charged particles that constitute a plasma.
Because charged particles roughly follow magnetic field lines, it is required for confinement that the magnetic field lies tangent to a foliation of nested surfaces.
Nested surfaces also ensure that large temperature and pressure gradients can be sustained in a steady state plasma, satisfying the necessary conditions for net fusion gain.

At large scales, the plasma state inside of a stellarator is often modeled using ideal magnetohydrodnamics (MHD), a set of moment equations for a single fluid coupled with Maxwell's equations.
At baseline, it is typically assumed that stellarators operate near \textit{MHD equilibrium}, obtained by setting the flow and time derivatives in ideal MHD to zero.
This results in deceptively simple equations for the magnetic field $\bm B$, where the magnetic Lorentz force balances with the fluid pressure $p$ via
\begin{align}
    \mu_0^{-1}(\nabla\times\bm{B})\times\bm{B} = \nabla p,\quad \nabla\cdot\bm{B} =0.\label{MHS}
\end{align}
MHD equilibria with suitable boundary conditions are the background for higher fidelity stability and transport models, including neoclassical transport and gyrokinetic turbulence. 

In principle, stellarator modeling requires self-consistently coupling small kinetic scales to the coarse-grained MHD scales, e.g.~through turbulent or neoclassical fluxes of heat and momentum. 
This extreme multiscale challenge motivates modern stellarator optimization studies to first translate small-scale design goals, such as single-particle confinement, into constraints on the MHD equilibrium model, and then search for MHD equilibria that approximately satisfy the constraints. 
More broadly, the global stellarator fusion program depends on the validity of MHD equilibrium at large scales to support downstream theory and computation.


The central importance of the MHD equilibrium model amplifies the significance of its shortcomings, which include the following.

\noindent \textbf{Challenges with mesh refinement:} In contrast to MHD equilibrium solvers for tokamaks, solvers in non-axisymmetric $3D$ geometries behave poorly under mesh refinement. 
Mainstream tools like VMEC \cite{hirshman_steepest-descent_1983} and DESC \cite{dudt_desc_2020,panici_desc_2023}, which assume nested flux surfaces, predict current sheets near rational surfaces with widths proportional to the radial mesh resolution \cite{lazerson_verification_2016}. These current sheets contradict the basic length scale assumptions of ideal MHD, including that the fluid length scale is significantly larger than the ion gyroradius.
A proposed solution to the apparent contradiction is to allow for the formation of islands and chaos, e.g.~using a so-called iterative solver like HINT2 \cite{suzuki_development_2006}, SIESTA \cite{hirshman_siesta_2011}, MRX \cite{blickhan_mrx_2025}, or PIES \cite{reiman_calculation_1986,drevlak_pies_2005}.
However, these solvers inexorably face mesh refinement issues as well. 
Ideal force balance requires constant plasma pressure along field lines. 
As soon as field-line integrability breaks even slightly, the field-line phase portrait fractures, producing a fractal mixture of integrability and chaos \cite{kraus_theory_2017}. 
The only smooth pressure profile constant along field lines in this scenario is globally constant. 
Hudson-Nakajima previously argued \cite{hudson_pressure_2010} that allowing for non-constant pressure in MHD equilibria with non-integrable magnetic fields leads to discontinuous pressure gradients and current densities. 
Although the solution strategy underlying SPEC \cite{hudson_computation_2012} isolates these discontinuities to a few ideal surfaces by appealing to a result due to Bruno-Laurence \cite{bruno_existence_1996}, it fails to avoid the smoothness issue altogether. 
Lack of smoothness under mesh refinement, both for iterative solvers and nested-surface solvers, fundamentally conflicts with MHD equilibria as coarse-grained plasma states.

\noindent \textbf{Degenerate potential energy:} The potential energy landscape for ideal MHD suffers from degeneracies that prevent energy minimization from selecting a unique state, even when the MHD equilibrium is stable according to the energy principle \cite{bernstein_energy_1958}.
We give a fully nonlinear demonstration of this degeneracy in Appendix \ref{sec:math_motivation} and further discussion in the context of second variation analysis in Section \ref{sec:improved_energy_landscape}. 
These analyses show that MHD equilibria can always be altered without changing potential energy by appropriately concentrating perturbations along rational surfaces. 
Such flat spots in the potential energy landscape create solution degeneracy inherent to the $3D$ MHD equilibrium problem, which in turn muddies the theoretical basis for iterative solvers. 
These solvers introduce numerical relaxation schemes that allow the formation of islands and chaos in the computed magnetic field. 
Although often physics-inspired, the relaxation schemes fall short of \emph{ab initio} reconnection models in order to reduce computational costs. 
Iterative solvers therefore break solution degeneracy implicitly, via specific details of numerical relaxation, without an entirely physical basis. 
While island locations likely enjoy robustness to relaxation method, at least in near-integrable fields, the robustness of island widths and internal structure is much less clear. 
Aside from computational challenges, the degenerate potential energy landscape also prevents application of the direct method in the calculus of variations to rigorously prove existence of MHD equilibria in $3D$. 
This provides support for H. Grad's famous conjecture \cite{grad_toroidal_1967} that asserts inherently three-dimensional MHD equilibria with nested flux surfaces cannot exist.

\noindent \textbf{Missing kinetic physics} Although MHD equilibrium calculations routinely influence neoclassical and gyrokinetic calculations by defining background field geometry, the MHD equilibrium model alone fails to account for feedback of kinetic scales on equilibrium scales. 
For example, the MHD equilibrium model ignores the bootstrap current, which potentially leads to overestimates of stellarator performance metrics, such as quality of quasisymmetry. 
To address this risk, various stellarator computational tools attempt to incorporate a self-consistent equilibrium bootstrap current by combining neoclassical transport and equilibrium calculations \cite{watanabe_three-dimensional_1992,landreman_optimization_2022}.
Although the theoretical justification for such iterations suffers from gaps (e.g. what precisely is the underlying monolithic equilibrium model that iteration converges upon?), their use in real stellarator design studies underscores the importance of the physics missing from the MHD equilibrium model.

Taken together, these shortcomings suggest that the large scale structure of a stellarator plasma may not be described by the MHD equilibrium model after all.

In this Article we present a new stellarator equilibrium model that improves upon each of the shortcomings inherent to the traditional MHD equilibrium model.
We dispense with the notion that stellarators satisfy static ideal MHD equilibrium conditions at large scales. 
Instead we assert that nonideal magnetic fluctuations---a piece of the physics missing from standard ideal MHD equilibria---significantly affect the solution near singular surfaces. 
Then we replace MHD equilibrium with \textit{statistical equilibrium}: large-scale forces only balance on average (Sec.~\ref{sec:derivation}). 
We derive the statistical equilibrium model by averaging Grad's ideal MHD potential energy functional, where the only assumptions are that the magnetic fluctuations are fast, ergodic, and small amplitude (there is no length scale assumption). 
We show analytically that statistical equilibria respond smoothly to small resonant perturbations in a simple slab geometry with a known length scale (Sec.~\ref{sec:asymptotics}). 
We demonstrate numerically that statistical equilibria continue to respond smoothly as the boundary perturbation amplitude increases into the nonlinear regime (Sec.~\ref{sec:nonlinear_computation}). 
In particular we show that the solver robustly converges under mesh refinement and predicts resolution-independent current sheet widths. 
We show that averaging regularizes the ideal MHD potential energy landscape by proving short-scale positive-definiteness for the second variation of averaged potential energy at statistical equilibrium (Sec.~\ref{sec:improved_energy_landscape}). 
Equivalently, we demonstrate ellipticity of the statistical equilibrium model, which stands in stark contrast to the mixed elliptic-hyperbolic type of the MHD equilibrium model. 
Finally, in Sec.~\ref{sec:discussion}, we summarize and discuss implications of the statistical equilibrium model for stellarator theory and design.


\section{Derivation of the statistical equilibrium model\label{sec:derivation}}

In the standard picture of stellarator plasmas, ideal MHD describes the largest scales. When expressed in Lagrangian variables, this model encodes the plasma configuration using a back-to-labels map $G$, which assigns a fluid parcel label $\bm{X} = G(\bm{x})\in Q_0$ to each point $\bm{x}$ in the fluid container $Q$. In other words, $G$ answers the question ``which particle is presently located at $\bm{x}$?" The Newcomb Lagrangian,
\begin{align*}
    L(G,\dot{G}) = \frac{1}{2}\int |\bm{u}(\bm{x})|^2 \rho(\bm{x}) \,d^3\bm{x}  - \int \mathcal{U}(\rho(\bm{x}),s(\bm{x}))\,\rho(\bm{x})\,d^3\bm{x} - \frac{1}{2}\mu_0^{-1}\int |\bm{B}(\bm{x})|^2\,d^3\bm{x},
\end{align*}
then governs the large-scale dynamics, where the equation of state is specified by the internal energy $\mathcal{U}(\rho,s)$. Without loss of generality, the reference space $Q_0$ and fluid container $Q$ can be taken as identical sets, with the same Riemannian metric. The fluid velocity $\bm{u}$, mass density $\rho$, entropy $s$, and magnetic field $\bm{B}$ are each expressed in terms of the back-to-labels map according to
\begin{gather}
    \bm{u}(\bm{x}) = -\mathbb{G}(\bm{x})^{-1}\dot{G}(\bm{x}),\nonumber\\
    \rho(\bm{x})  = \mathcal{J}(\bm{x})\,\rho_0(G(\bm{x}))\,,\quad s(\bm{x}) = s_0(G(\bm{x})),\quad \bm{B}(\bm{x}) = \mathbb{A}(\bm{x})\bm{B}_0(G(\bm{x})), \label{ideal_advection}\\
    \mathbb{G}(\bm{x}) = DG(\bm{x}),\quad \mathbb{A}(\bm{x}) = \text{adj}\,\mathbb{G}(\bm{x}),\quad \mathcal{J}(\bm{x}) = \text{det}\,\mathbb{G}(\bm{x}).\nonumber
\end{gather}
The columns of $\mathbb{G}$ give the partial derivatives of $G$ and $\mathbb{A}$ denotes the adjugate of $\mathbb{G}$. The reference-space fields $\rho_0$, $s_0$, and $\bm{B}_0$ denote the plasma mass density, entropy, and magnetic field in some reference configuration of fluid parcels. The formulas \eqref{ideal_advection} express advection of mass, entropy, and magnetic flux by the flow. In static equilibrium, $\bm{u} = 0$ and the plasma configuration $G$ is a critical point for the potential energy $\mathcal{W}(G)$:
\begin{align}
    \mathcal{W}(G) & =  \int \mathcal{U}(\rho(\bm{x}),s(\bm{x}))\,\rho(\bm{x})\,d^3\bm{x} + \frac{1}{2}\mu_0^{-1}\int |\bm{B}(\bm{x})|^2\,d^3\bm{x}.\label{newcomb_functional}
\end{align}
This, together with the physical requirement $\nabla\cdot\bm{B}_0 = 0$, implies the equations \eqref{MHS} defining MHD equilibrium.

Due to the documented spurious small scale structures in three-dimensional MHD equilibrium computations (cf Section \ref{introduction}), we reject the notion that MHD equilibrium describes the large-scale structure of stellarator plasmas in steady-state. To improve the steady-state model, we introduce the following physically-motivated hypotheses.
\begin{enumerate}
    \item[(H1)] The large-scale plasma configuration is given by a back-to-labels map $G$, as in the standard model. The plasma potential energy is still given by Eq.\,\eqref{newcomb_functional}.
    \item[(H2)] The reference magnetic field $\bm{B}_0 = \mathcal{J}^{-1}\mathbb{G}\bm{B}$ sustains fluctuations due to nonideal evolution of $\bm{B}$. The fluctuations vary on a timescale much shorter than the evolution timescale for $G$. We do not assume any space scale separation. The reference thermodynamic variables $\rho_0,s_0$ remain constant in time, as in ideal MHD. Velocity fluctuations are neglibile.
    \item[(H3)] In steady-state $\dot{G} = 0$. The plasma need not achieve instantaneous force balance. However, forces balance when averaged over the short fluctuation timescale.
    \item[(H4)] Fluctuations are ergodic. There is a $\sigma$-algebra $\mathcal{F}_0$ and probability measure $P_0$ on the space $\mathcal{B}_0$ of reference magnetic fields such that time averaging over the short fluctuation timescale is equivalent to ensemble averaging with respect to the probability space $(\mathcal{B}_0,\mathcal{F}_0,P_0)$.
\end{enumerate}

Granting these hypotheses, the following model for large-scale structure of a steady-state stellarator plasma emerges. We refer to this model as \textbf{statistical equilibrium}. Suppose a stellarator has reached steady-state. By hypothesis (H1) and negligibility of velocity fluctuations from (H2), the instantaneous force on the plasma is given by first variation of the potential energy \eqref{newcomb_functional} with respect to $G$. By (H3), the first variation of $\mathcal{W}$ at $G$ therefore vanishes after time averaging over the fluctuation timescale $T$,
\begin{align*}
    \frac{1}{T}\int_{t-T/2}^{t+T/2}\delta \mathcal{W}(G)\,d\overline{t} = 0.
\end{align*}
By (H2) and (H3), the time dependence in $\delta \mathcal{W}(G)$ is caused only by fluctuations in $\bm{B}_0$. Finally, by (H4), we find
\begin{align}
    \delta \Wbar(G) = 0,
\end{align}
where $\Wbar = \mathbb{E}[\mathcal{W}]$ denotes the mean potential energy computed relative to the probability measure $P_0$. The mean potential energy is given explicitly in the label quanties by 
\begin{align}
    \Wbar(G) &= \frac{1}{2}\mu_0^{-1}\mathbb{E}\bigg[\int\text{tr}\bigg( \bm{B}_0(G(\bm{x}))\bm{B}_0(G(\bm{x}))^T\mathbb{A}(\bm{x})^T\mathbb{A}(\bm{x})\bigg)\,d^3\bm{x}\bigg]+\int \mathcal{U}(\rho(\bm{x}),s(\bm{x}))\,\rho(\bm{x})\,d^3\bm{x}\nonumber\\
    & =  \frac{1}{2}\mu_0^{-1}\int\text{tr}\bigg( \bm{\mathsf{B}}_0(G(\bm{x}))\bm{\mathsf{B}}_0(G(\bm{x}))^T\mathbb{A}(\bm{x})^T\mathbb{A}(\bm{x})\bigg)\,d^3\bm{x}+\int \mathcal{U}(\rho(\bm{x}),s(\bm{x}))\,\rho(\bm{x})\,d^3\bm{x}\nonumber\\
    & +  \frac{1}{2}\mu_0^{-1}\int\text{tr}\bigg( \Sigma_0(G(\bm{x}))\mathbb{A}(\bm{x})^T\mathbb{A}(\bm{x})\bigg)\,d^3\bm{x},\nonumber
\end{align}
where we have defined the reference mean and variance of the magnetic field as
\begin{gather*}
   \Bobar(\bm{X}) = \mathbb{E}[\bm{B}_0(\bm{X})],\quad \Sigma_0(\bm{X}) = \mathbb{E}\bigg[[\bm{B}_0(\bm{X})-\Bobar(\bm{X})][\bm{B}_0(\bm{X}) - \Bobar(\bm{X})]^T\bigg].
\end{gather*}
Equivalently, in the spatial frame, we have
\begin{gather}
\label{mean_potential_energy}
    \Wbar(G) =  \frac{1}{2}\mu_0^{-1}\int|\Bbar(\bm{x})|^2\,d^3\bm{x}+\int \mathcal{U}(\rho(\bm{x}),s(\bm{x}))\,\rho(\bm{x})\,d^3\bm{x}+  \frac{1}{2}\mu_0^{-1}\int\text{tr}\,\Sigma(\bm{x})\,d^3\bm{x},\\
\nonumber
    \Bbar(\bm{x}) = \mathbb{A}(\bm{x})\Bobar(G(\bm{x})), \quad \Sigma(\bm{x}) = \mathbb{A}(\bm{x})\Sigma_0(G(\bm{x}))\mathbb{A}^T(\bm{x}).
\end{gather}
Note that the first two terms in \eqref{mean_potential_energy} sum to give the usual potential energy, but with the mean reference magnetic field replacing the reference magnetic field. In summary, the statistical equilibrium model asserts that the large-scale plasma configuration $G$ is a critical point for the mean potential energy \eqref{mean_potential_energy}.

The strong-form Euler-Lagrange equations for $\Wbar$ may be computed using the Euler-Poincar\'e formalism \cite{holm_eulerpoincare_1998} as follows. Assume $G:Q\rightarrow Q_0$ is a diffeomorphism, i.e. smooth and smoothly invertible. If $\delta G$ denotes an infinitesimal variation of $G$, let $\bm{\xi}(\bm{x}) = -\mathbb{G}(\bm{x})\delta G(\bm{x})$ denote its Eulerianization. The variations in $\bm{B}$, $\rho$, and $s$ induced by $\delta G$ are given by
\begin{align*}
    \delta \bm{B} = \nabla\times(\bm{\xi}\times\bm{B}),\quad \delta\rho = -\nabla\cdot(\bm{\xi}\rho),\quad \delta s = -\bm{\xi}\cdot\nabla s.
\end{align*}
The first variation of $\Wbar$ at $G$ in the direction $\delta G$ is therefore
\begin{align*}
    \delta\Wbar(G)[\delta G] &= \mathbb E[\delta \mathcal W[\delta G]] \\
    & = \mu_0^{-1}\mathbb{E}\left[\int \bigg(\nabla\cdot\left[\frac{1}{2}|\bm{B}|^2\mathbb{I} - \bm{B}\bm{B}^T\right]\bigg)\cdot\bm{\xi}\,d^3\bm{x}\right]+ \int \nabla p\cdot\bm{\xi}\,d^3\bm{x}\\
    & = \int \nabla\cdot \bigg(p\mathbb{I}+\mu_0^{-1}\frac{1}{2}|\Bbar|^2\mathbb{I} - \mu_0^{-1}\Bbar\Bbar^T + \frac{1}{2}\mu_0^{-1}\text{tr}(\Sigma)\mathbb{I} - \mu_0^{-1}\Sigma\bigg)\cdot\bm{\xi}\,d^3\bm{x}.
\end{align*}
Here we have used $\bm{\xi} = 0$ on $\partial Q$ for integration by parts, which is implied by the diffeomorphism property for $G$, to eliminate several boundary terms. We have also used the standard thermodynamic definition of pressure $p = \rho^2\partial_\rho\mathcal{U}$. Since $\bm{\xi}$ is arbitrary away from $\partial Q$, the strong-form Euler-Lagrange equations are
\begin{align}
    \nabla\cdot \bigg(p\mathbb{I}+\mu_0^{-1}\frac{1}{2}|\Bbar|^2\mathbb{I} - \mu_0^{-1}\Bbar\Bbar^T + \frac{1}{2}\mu_0^{-1}\text{tr}(\Sigma)\mathbb{I} - \mu_0^{-1}\Sigma\bigg)=0.\label{statistical_equilibrium_divergence_form}
\end{align}
When $\Sigma_0 = 0$, these equations reduce to the usual MHD equilibrium model written in divergence form. It follows that statistical equilibrium modifies MHD equilibrium by adding a self-consistent magnetic Reynolds stress to the total stress tensor.

In principle, the statistical equilibrium model allows for any choice of domain $Q$, any equation of state $\mathcal{U}(\rho,s)$, any reference thermodynamic fields $\rho_0,s_0$, and any probability measure $P_0$. In particular, the mean reference magnetic field need not have an integrable phase portrait. In this work we restrict attention to the following simple set of choices for these modeling parameters, leaving the implications of other choices to future investigation. We make these choices in order to minimize deviation between statistical equilibrium modeling and the most broadly adopted modeling paradigm implemented in codes like VMEC and DESC.

\textbf{Fluid container and reference space:} The fluid container $Q$ and the reference domain $Q_0$ are each diffeomorphic to $I\times \mathbb{T}^2$, where $I\subset \mathbb{R}$ denotes a closed interval and $\mathbb{T}^2 = S^1\times S^1 = (\mathbb R\backslash 2\pi\mathbb Z)^2$ denotes the $2$-torus. We choose coordinates $(V,\Theta,Z)\in I\times \mathbb{T}^2$ on $Q_0$ such that the reference volume form is given by
\begin{align*}
    d^3\bm{X} = \frac{1}{(2\pi)^2}dV\,d\Theta\,dZ,
\end{align*}
and the reference domain boundaries are given by $V=0$ and $V=L_0^3$. Here $L_0^3$ denotes the fluid container volume.

\textbf{Equation of state:} The equation of state is
\begin{align*}
    \mathcal{U}(\rho,s) = -c_0\exp(s/c_V)\rho^{-1},
\end{align*}
where $c_0$ is a positive constant. This corresponds to the unphysical scenario of an ideal gas with vanishing ratio of specific heats $\gamma = 0$. In the MHD equilibrium model the value of $\gamma$ does not change the set of solutions with $\nabla p$ nowhere-vanishing. It is a common practice, going back to H. Grad \cite{grad_hydromagnetic_1958,grad_new_1964}, to exploit this formal flexibility in order to simplify the potential energy functional used to compute equilibria. This explains our choice. However, it is important to recognize that the value of $\gamma$ may affect the set of solutions of the statistical equilibrium model, even though this does not happen for MHD equilibria. Since pressure $p = c_0\,\exp(s/c_V)$ only depends on entropy for this equation of state, we follow tradition by exchanging $s$ and $p$. 

\textbf{Reference thermodynamic fields:} The deterministic reference pressure is taken as a function of $V$ only
\begin{align*}
    p_0(V,\Theta,Z) = \mathsf{p}(V).
\end{align*}
The reference mass density $\rho_0$ is irrelevant because it does not appear in the mean potential energy $\Wbar$.

\textbf{Reference magnetic field ensemble:} The random reference magnetic field is given by 
\begin{align}
        \bm{B}_0 = \psi_T'(V)\,\nabla V\times\nabla\Theta - \psi_P'(V)\,\nabla V\times \nabla Z,\label{simple_fluctuation_model}
\end{align}
where $\psi_T,\psi_P$ are random single-variable profile functions. Note that any realization of $\bm{B}_0$ has an integrable phase portrait, with a foliation by nested flux surfaces. The mean profiles are denoted
\begin{align*}
    \Psi_T(V) = \mathbb{E}[\psi_T(V)],\quad\Psi_P(V) = \mathbb{E}[\psi_P(V)].
\end{align*}
To specify $\Sigma_0$ we impose isotropic second-order statistics on the differentiated profiles according to
\begin{gather*}
    \mathbb{E}\bigg[[\psi_P'(V) - \Psi_P'(V)]^2\bigg] =\mathbb{E}\bigg[[\psi_T'(V) - \Psi_T'(V)]^2\bigg]= \frac{\delta\Psi_0^2}{\ell_0^6}\\
    \mathbb{E}\bigg[[\psi_P'(V) - \Psi_P'(V)][\psi_T'(V) - \Psi_T'(V)]\bigg]  = 0.
\end{gather*}
Here $\delta \Psi_0$ and $\ell_0$ denote a characteristic amplitude and length scale for nonideal fluctuations in magnetic flux, with the size of magnetic fluctuations going as $\delta B_0 = \ell_0^{-3} L_0 \delta \Psi_0$. The mean reference magnetic field and the reference variance $\Sigma_0$ are then
\begin{align}
    \Bobar(\bm{X}) & = \Psi_T'(V)\nabla V\times\nabla\Theta - \Psi_P'(V)\nabla V\times\nabla Z\label{mean_reference_field}\\
    \Sigma_0(\bm{X}) &= \frac{\delta\Psi_0^2}{\ell_0^6}(\nabla V\times \nabla \Theta)(\nabla V\times \nabla \Theta)^T + \frac{\delta\Psi_0^2}{\ell_0^6}(\nabla V\times \nabla Z)(\nabla V\times \nabla Z)^T.\label{reference_variance}
\end{align}

These modeling choices imply simplified expressions for the mean potential energy and its associated Euler-Lagrange equations. Denote the components of $G$ as $G = (\rad,\theta,\zeta)^T$. The mean potential energy is
\begin{align}
    \Wbar(G) & = \frac{1}{2}\mu_0^{-1}\int |\Psi_T'(\rad)\nabla \rad\times \nabla\theta - \Psi_P'(\rad)\nabla \rad\times\nabla \zeta|^2\,d^3\bm{x} -\int \mathsf{p}(\rad)\,d^3\bm{x}\nonumber\\
    &+ \frac{1}{2}\mu_0^{-1}\frac{\delta\Psi_0^2}{\ell_0^6}\int \bigg(|\nabla \rad\times\nabla\theta|^2 + |\nabla \rad\times\nabla\zeta|^2\bigg)\,d^3\bm{x}.\label{mean_potential_energy_simplified}
\end{align}
The associated strong-form Euler-Lagrange equations are equivalent to
\begin{align}
    &0=-\nabla\cdot(\overline{\mu} \nabla \rad) + \mu_0^{-1}(\Psi_T''\nabla\theta - \Psi_P''\nabla\zeta)^T \nabla \rad_\times^T\nabla\rad_\times (\Psi_T'\nabla\theta - \Psi_P'\nabla\zeta) -\mathsf{p}'\label{statistical_equilibrium_v}\\
    &0=\nabla\cdot(\nabla v_\times^T\nabla v_\times \nabla\theta)\label{statistical_equilibrium_theta}\\
    &0 = \nabla\cdot(\nabla v_\times^T\nabla v_\times \nabla\zeta),\label{statistical_equilibrium_zeta}
\end{align}
where $\bm{v}_\times$ denotes the $3\times 3$ matrix defined by $\bm{v}_\times\bm{w} =\bm{v}\times\bm{w}$ and we have introduced the positive-definite matrix
\begin{align*}
    \overline{\mu} = \mu_0^{-1}(\Psi_T'\nabla\theta - \Psi_P'\nabla\zeta)_\times^T(\Psi_T'\nabla\theta - \Psi_P'\nabla\zeta)_\times + \mu_0^{-1}\frac{\delta\Psi_0^2}{\ell_0^6}\nabla\theta_\times^T\nabla\theta_\times + \mu_0^{-1}\frac{\delta\Psi_0^2}{\ell_0^6}\nabla\zeta_\times^T\nabla\zeta_\times.
\end{align*}
Note that the functional $\Wbar(G)$ is invariant under the family of symmetries $\theta\mapsto \theta + h(v),$ $\zeta\mapsto \zeta + g(v)$, where $h,g : I \to \mathbb{R}$ are arbitrary smooth single-variable functions. When the fields $\theta,\zeta$ are viewed as coordinates on the $v$-surfaces, these transformations correspond to independently shifting the origin on each $v$-surface. These transformations are remnants of the Grad functional's symmetries that allow for arbitrary specification of the toroidal angle in equilibrium solvers like VMEC. Due to the presence of these symmetries, when solving the Euler-Lagrange equations it will always be necessary to fix the origin of each $v$-surface by introducing some convenient rule. For example, after introducing a fixed curve that transversally intersects the $v$-surfaces, one viable rule would set $\theta=0$ and $\zeta=0$ along the curve. This is not the only viable choice.

\section{Asymptotic response to boundary perturbations}
\label{sec:asymptotics}
This Section uses asymptotic matching theory to formally demonstrate two remarkable properties of statistical equilibrium: (1) away from rational surfaces, statistical equilibria nearly coincide with MHD equilibria; (2) near rational surfaces, statistical equilibria are smooth while MHD equilibria are singular. Taken together, these properties suggest viability of statistical equilibrium as a large-scale model for stellarator plasmas, in contrast to the MHD equilibrium model. 

To clarify our analysis, in this and all following Sections we adopt dimensionless variables. Introduce a characteristic equilibrium magnetic flux $\Psi_0$ and a characteristic plasma pressure $p_0$. Measure lengths in units of the fluid container scale length $L_0$, magnetic flux in units of $\Psi_0$ (and therefore the magnetic field in units of $B_0 = L_0^{-2} \Psi_0$), and pressure in units of $p_0$. The dimensionless statistical equilibrium equations are then those given in Section \ref{sec:derivation} with the substitutions 
\begin{align}
    \mu_0^{-1}\rightarrow 1,\quad \mathsf{p}\rightarrow \beta\,\mathsf{p},\quad \mu_0^{-1}\frac{\delta\Psi_0^2}{\ell_0^6}\rightarrow \lambda^2.\label{substitutions}
\end{align}
Here
\begin{align}
    \beta = \frac{p_0\,L_0^4}{\mu_0^{-1}\Psi_0^2},\quad \lambda^2 = \frac{L_0^6}{\ell_0^6}\frac{\delta\Psi_0^2}{\Psi_0^2} = \bigg(\frac{\delta B_0}{B_0}\bigg)^2,\label{lambda_def}
\end{align}
are dimensionless parameters that should each be small in a stellarator plasma. The first parameter $\beta$ denotes the ratio of thermal energy to magnetic energy. It quantifies efficiency of the magnetic confinement system. The second parameter $\lambda^2$ represents a normalized variance of nonideal fluctuations in the reference magnetic field. It quantifies the statistical deviation from MHD equilibrium. While the precise value of $\lambda$ is unknown, magnetic fluctuations in W7-X have been observed with values between $\lambda = 10^{-3}$ through $\lambda=10^{-5}$ \cite{rahbarnia_alfvenic_2020}.

\begin{figure}
    \centering
    \includegraphics[width=0.75\linewidth]{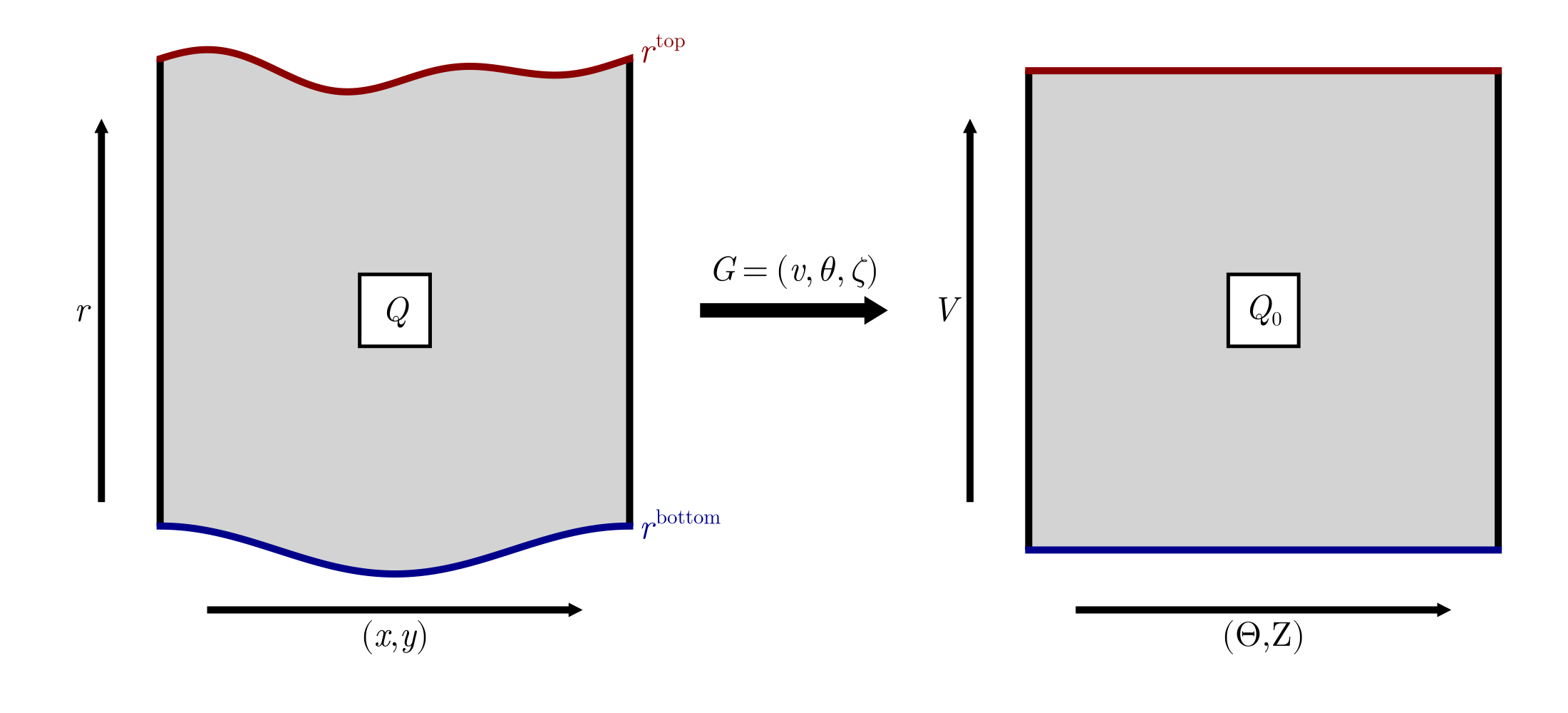}
    \caption{A schematic of the considered geometry. On the left, the fluid domain $Q$ is bounded by $r^{\text{bottom}}$ and $r^{\text{top}}$ in dark blue and red respectively. The back-to-labels map $G$ maps $Q$ to the reference domain $Q_0$, with the boundary condition that the top and bottom fluid boundaries must map to the corresponding top and bottom boundaries in the reference configuration.}
    \label{fig:domains}
\end{figure}

We seek solutions of the statistical equilibrium model when the fluid container $Q$ is flat (in the sense of Riemannian geometry) with small-amplitude boundary perturbations. See Fig. \ref{fig:domains}. Equip $Q$ with a fixed coordinate system $(r,x,y)\in \mathbb{R}\times \mathbb{T}^2$, where $r$ denotes a radial coordinate, $x\in S^1$ denotes a poloidal angle, and $y\in S^1$ denotes a toroidal angle. The metric tensor is $g = dr^2 + dx^2 + dy^2$. The ``top" of the domain is given by the graph $r  = r^{\text{top}}(x,y)$, while the ``bottom" of the domain is given by $r = r^{\text{bottom}}(x,y)$. 
In this section, we assume $r^{\text{top}}$ and $r^{\text{bottom}}$ are each smooth functions that depend on a small parameter $\epsilon$ proportional to the amplitude of the boundary perturbations. For concreteness we set $r^{\text{top}}(x,y) = 1 + \epsilon\,r_1^{\text{top}}(x,y)$ and $r^{\text{bottom}} = \epsilon\,r_1^{\text{bottom}}(x,y)$, which ensures that the boundary components of the domain are each flat when $\epsilon = 0$. The space domain $Q$ is
\begin{align*}
    Q = \{(r,x,y)\in \mathbb{R}\times\mathbb{T}^2\mid r^{\text{bottom}}(x,y)\leq r\leq r^{\text{top}}(x,y) \}.
\end{align*}
We continue to use the coordinates $(V,\Theta,Z)\in [0,1]\times\mathbb{T}^2$ on the reference domain $Q_0$. The boundary conditions on the back-to-labels map $G = (v,\theta,\zeta)^T$ are therefore
\begin{align}
    v(r^{\text{top}}(x,y),x,y) = 1,\quad  v(r^{\text{bottom}}(x,y),x,y) = 0.\label{fundamental_bc}
\end{align}
The three components of $G$ define the dependent variables for the statistical model. These component functions must satisfy the dimensionless versions of Eqs.\,\eqref{statistical_equilibrium_v}-\eqref{statistical_equilibrium_zeta}, found most easily using the substitution rule \eqref{substitutions}. 
We would like to understand the behavior of solutions of these equations for small $\epsilon$ and $\lambda$. To that end we expand the solution in powers of $\epsilon$ first, before studying small-$\lambda$ using a subsidiary expansion.

The unperturbed ($\epsilon=0$) solution is $G_0(r,x,y) = (v_0(r),x,y)^T$. The function $v_0$ satisfies a first-order ordinary differential equation that expresses total pressure balance, including magnetic Reynolds stress,
\begin{gather}
    \frac{1}{2}\bigg(\partial_r\psi_T\bigg)^2 +\frac{1}{2}\bigg(\partial_r\psi_P\bigg)^2  +\beta\,\mathsf{p}(v_0) + \lambda^2(\partial_r v_0)^2 = \Pi_0,\label{leading_order_pressure_balance}\\
    \quad \psi_T(r) = \Psi_T(v_0(r)),\quad\psi_P(r) = \Psi_P(v_0(r)).\nonumber
\end{gather}
The constant $\Pi_0$ denotes the total unperturbed pressure. The solution $v_0$ must satisfy the boundary conditions $v_0(0) = 0$ and $v_0(1) = 1$. The lower boundary condition $v_0(0) = 0$ determines the integration constant for the first-order equation \eqref{leading_order_pressure_balance}. The constant $\Pi_0$ must be adjusted appropriately to satisfy the upper boundary condition  $v_0(1) = 1$.

Now consider a perturbed solution $G_\epsilon = (v_\epsilon,\theta_\epsilon,\zeta_\epsilon)$ of the form
\begin{align*}
    v_\epsilon(r,x,y) & = v_0(\chi_\epsilon(r,x,y))\\
    \chi_\epsilon(r,x,y) &= r - \epsilon\,R_1(r,x,y) + \epsilon^2\,R_2(r,x,y) + \dots\\
    \theta_\epsilon(r,x,y)& = x - \epsilon\,X_1(r,x,y) + \epsilon^2\,X_2(r,x,y)+\dots\\
    \zeta_\epsilon(r,x,y) & = y - \epsilon\,Y_1(r,x,y) + \epsilon^2\,Y_2(r,x,y)+\dots.
\end{align*}
Here all coefficient functions are single-valued functions on $Q$. We will attempt to solve for the first-order solution $\bm{\eta} = (R_1,X_1,Y_1)$.

The variational structure of statistical equilibrium implies $\bm{\eta}$ is a critical point for the second variation of the mean potential energy $\Wbar$ at $G_0$. A straightforward calculation reveals the following explicit formula for the second variation, modified appropriately to allow for complex-valued $\bm{\eta}$:
\begin{align*}
    \delta^2\Wbar(G_0)[\bm{\eta}^*,\bm{\eta}] & =\int \frac{|\Bbar\times(\Bbar\cdot\nabla\bm{\eta})|^2}{|\bm{B}|^2}\,d^3\bm{x} + \int \left|\nabla\cdot\bm{\eta} -\frac{\Bbar\cdot\nabla\bm{\eta}\cdot\Bbar}{|\Bbar|^2}\right|^2\,|\Bbar|^2\,d^3\bm{x}\nonumber\\
    & + \lambda^2\int \frac{|\bm{e}_T\times(\bm{e}_T\cdot\nabla\bm{\eta})|^2}{|\bm{e}_T|^2}\,d^3\bm{x} +\lambda^2 \int \left|\nabla\cdot\bm{\eta} -\frac{\bm{e}_T\cdot\nabla\bm{\eta}\cdot\bm{e}_T}{|\bm{e}_T|^2}\right|^2\,|\bm{e}_T|^2\,d^3\bm{x}\nonumber\\
    & + \lambda^2\int \frac{|\bm{e}_P\times(\bm{e}_P\cdot\nabla\bm{\eta})|^2}{|\bm{e}_P|^2}\,d^3\bm{x} +\lambda^2 \int \left|\nabla\cdot\bm{\eta} -\frac{\bm{e}_P\cdot\nabla\bm{\eta}\cdot\bm{e}_P}{|\bm{e}_P|^2}\right|^2\,|\bm{e}_P|^2\,d^3\bm{x}.
\end{align*}
Here $\bm{e}_T = \nabla(v_0(r))\times\nabla x$, $\bm{e}_P = -\nabla(v_0(r))\times \nabla y$, and $\Bbar = \Psi_T'(v_0(r))\bm{e}_T + \Psi_P'(v_0(r))\bm{e}_P$. The essential boundary condition on $\bm{\eta}$ implied by \eqref{fundamental_bc} is
\begin{align}
    R_1(1,x,y) = r_1^{\text{top}}(x,y) ,\quad R_1(0,x,y) = r_1^{\text{bottom}}(x,y).\label{linearized_bcs}
\end{align}
The first-order solution $\bm{\eta}$ is governed by the Euler-Lagrange equations associated with the second variation, which agree with the linearization of the statistical equilibrium model about the unperturbed solution.

By homogeneity of the unperturbed solution in $(x,y)$, when solving the linearized equations it is sufficient to seek solutions of the form $\bm{\eta} = \overline{\bm{\eta}}(r)\exp(imx + iny)$, where $\overline{\bm{\eta}} = (\overline{R},\overline{X},\overline{Y})^T$ denotes an $r$-dependent vector of Fourier coefficients. Substituting this ansatz into the second variation and varying with respect to $\overline{X}$ and $\overline{Y}$ leads to a simple formula relating the angular variables to the radial variable when $n^2+m^2\neq 0$,
\begin{align}
    \overline{X} = \frac{im}{n^2 + m^2}\overline{R}',\quad \overline{Y} = \frac{in }{n^2+m^2}\overline{R}'.\label{angular_solution}
\end{align}
When $n=m=0$ we instead impose $\overline{X} = \overline{Y} = 0$, corresponding to fixing the origin on each flux surface. In either case, the angular solution can be substituted back into the full second variation $\delta^2\Wbar(G_0)[\bm{\eta}^*,\bm{\eta}]$ to find a reduced second variation $\delta^2\mathsf{w}(G_0)[\overline{R}^*,\overline{R}]$ involving only $\overline{R}$. This results in a reduction of the full nonlinear statistical equilibrium equations to a single scalar equation described in Section \ref{sec:improved_energy_landscape}. 
After various cancellations, the result is
\begin{align}
    \delta^2\mathsf{w}(G_0)[\overline{R}^*,\overline{R}] = \int D_{mn}(r)\,\bigg(|\overline{R}'|^2 + (n^2+m^2)|\overline{R}|^2\bigg)\,d^3\bm{x},\label{reduced_second_variation_slab}
\end{align}
where
\begin{align*}
    D_{mn}(r) = \begin{cases}(g_{mn}(r))^2 +\lambda^2(v_0'(r))^2& n^2+m^2\neq 0\\ (\partial_r\psi_T)^2 +(\partial_r\psi_P)^2 + 2\lambda^2 (v_0'(r))^2 & n^2+m^2 = 0\end{cases}.
\end{align*}
Here, the function $g_{mn}$ is the resonance defect
\begin{equation*}
    g_{mn}(r) = \frac{n}{\sqrt{n^2+m^2}}\partial_r\psi_T + \frac{m}{\sqrt{n^2+m^2}}\partial_r\psi_P.
\end{equation*}
The appropriate essential boundary conditions for $\overline{R}$ follow from Fourier decomposition of \eqref{linearized_bcs}:
\begin{align}
    \overline{R}(1) = r_{m,n}^{\text{top}} ,\quad \overline{R}_1(0) = r_{m,n}^{\text{bottom}},\label{fourier_bcs}
\end{align}
where $r_{m,n}^{\text{top}}$, $r_{m,n}^{\text{bottom}}$, denote the Fourier coefficients of the first-order boundary perturbations. We find that the strong-form Euler-Lagrange equations for $\overline{R}$ are given by
\begin{align}
    -\partial_r(D_{mn}(r)\partial_r\overline{R}) + (n^2+m^2)D_{mn}(r)\,\overline{R} = 0. \label{linearized_gse_fluct}
\end{align}
If these equations can be solved for $\overline{R}$ then the full first-order change in the solution $\bm{\eta}$ can be recovered using 

Assuming $v_0'$ is bounded away from zero, Equation \eqref{linearized_gse_fluct} is uniformly elliptic because $D_{mn}(r)$ is nowhere-vanishing. But as $\lambda\rightarrow 0$, the statistical equilibrium model approaches the non-elliptic MHD equilibrium model, which manifests as a boundary layer solution of Eq.\,\eqref{linearized_gse_fluct} concentrated at the rational surface $r=r_s$ defined by $g_{mn}(r_s) = 0$.  Asymptotic matching leads to a detailed picture of the statistical equilibrium solution near the rational surface for small $\lambda$, which we now treat as a subsidiary ordering parameter.

Introduce the scaled radial coordinate $q = (r-r_s)/\lambda$ and consider the corresponding inner layer equation
\begin{align*}
    \partial_q\left[\big((\partial_r g_{mn}(r_s))^2\,q^2 + v_0'^2(r_s)\big)\partial_q\overline{R}\right] = 0.
\end{align*}
The general solution of this equation is
\begin{align}
    \overline{R}(q) = \frac{1}{2}(R_+ + R_-) + (R_+ - R_-)\,\pi^{-1}\tan^{-1}\bigg(
    \frac{q}{L_{mn}(r_s)}\bigg), \label{inner_solution} 
\end{align}
where
\begin{equation}
    L_{mn}(r_s) = \abs{\frac{v_0'(r_s)}{\partial_r g_{mn}(r_s)}},\label{eq:layer-width}
\end{equation}
and the values $R_+$ and $R_-$ in Eq.~\ref{inner_solution} denote the limiting values for $\overline{R}(X)$ as $q\rightarrow \infty$ and $q\rightarrow -\infty$, respectively.
The formula \eqref{inner_solution} suggests how the statistical equilibrium model regularizes the singular behavior of MHD equilibrium near rational surfaces. Instead of producing true discontinuities, the new model predicts a smooth variation across the rational surface over a length scale set by the variance $\lambda^2$ of the profiles. The dimensional smoothing length is $\Lambda = \lambda L_0$. We interpret Eq.~\eqref{eq:layer-width} as the contribution to the layer width from the inverse magnetic shear, where we note that chain rule gives $L_{mn}=|\partial_{v_0} g_{mn}|^{-1}$, the derivative of the resonance condition with respect to the leading order flux variable. 

To determine the behavior of the solution away from the rational surface consider the outer layer equation obtained by setting $\lambda = 0$ in Eq.\,\eqref{linearized_gse_fluct}. Since this equation equation coincides with the $\overline{R}$-equation in MHD equilibrium, we infer that statistical equilibrium agrees with MHD equilibrium to first-order in $\lambda$ away from rational surfaces.
This equation has two independent solutions, one of which is singular at the rational surface and therefore incompatible with the inner layer solution. Let $R^*$ denote the other solution, specified by requiring $R^*=-1$ at the rational surface, where $\partial_r R^* = 0$ is required for consistency at the singularity. 
The outer solution to the left and right of the rational surface is given to leading order in $\lambda$ by
\begin{align*}
    R(r) =\begin{cases}
    R_L(r),\quad r < r_s,\\
    R_R(r),\quad r > r_s,
    \end{cases}
\end{align*}
where
\begin{equation*}
    R_L(r)= R(0)\frac{R^{*}(r)}{R^{*}(0)}, \qquad R_R(r)=R(1)\frac{R^{*}(r)}{R^{*}(1)}. 
\end{equation*}
Numerical simulations, as well as exact solutions of the outer equations with specific $\Psi_T,\Psi_P$ profiles, indicate $R^{*}(0) \neq 0$ and $R^{*}(1) \neq 0$.

\begin{figure}
    \centering
    \includegraphics[width=0.5\linewidth]{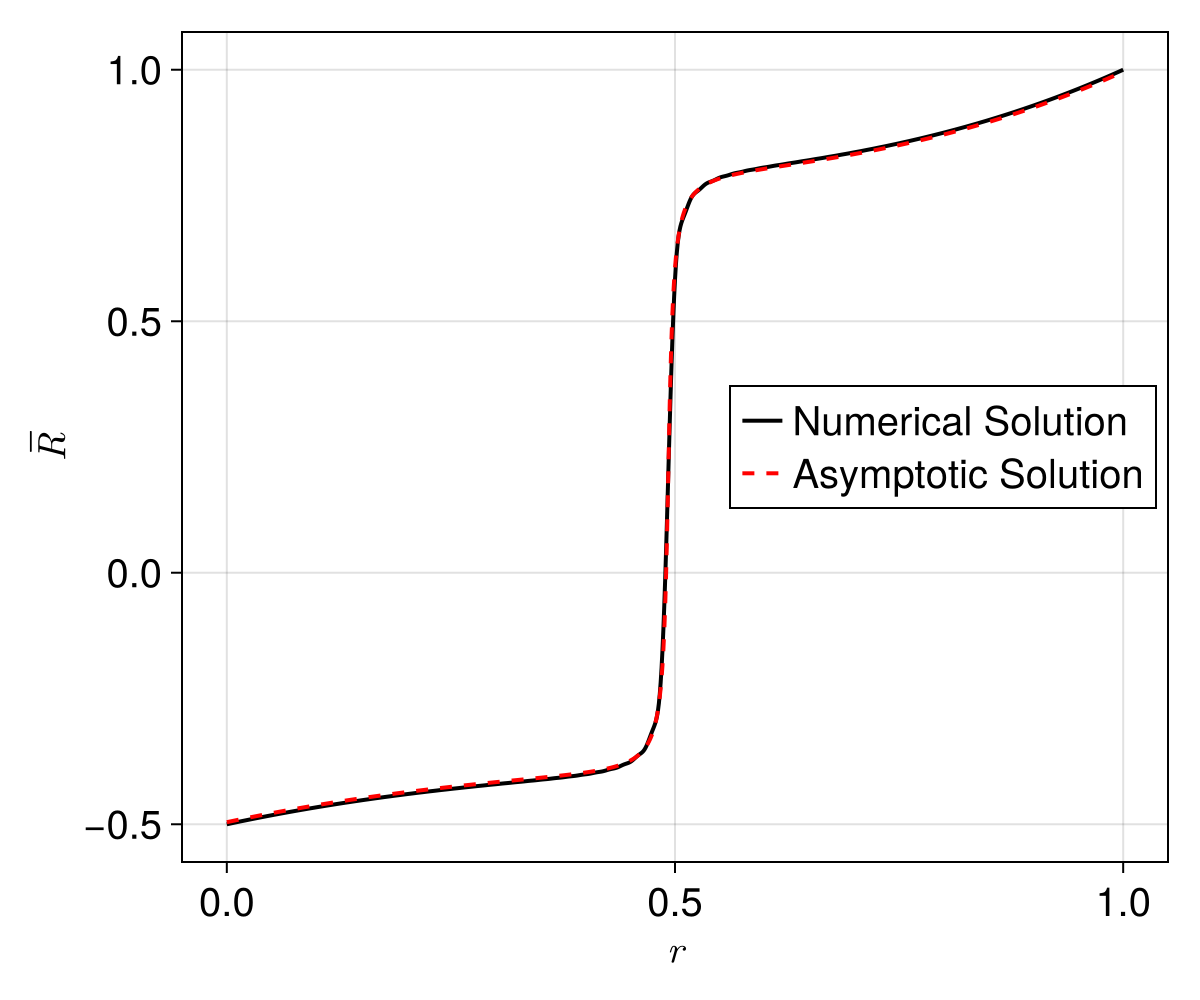}
    \caption{Comparison of a direct numerical solution of Eq.~\eqref{reduced_second_variation_slab} (black line) and the asymptotic solution \eqref{uniform_soln_second_order} (red dashed line).}
    \label{fig:asymptotic-solution}
\end{figure}

It is simple to verify that
\begin{align}
    \overline{R}(r) = \frac{1}{2}(R_R(r) + R_L(r)) + \frac{1}{\pi}(R_R(r) - R_L(r)) \tan^{-1} \bigg(\frac{r-r_s}{\lambda L_{mn}(r_s)} \bigg),\label{uniform_soln_second_order}
\end{align}
is an asymptotic solution for $\overline{R}(r)$, uniformly valid for $r\in[0,1]$.
In Fig.~\ref{fig:asymptotic-solution}, we test the asymptotic solution with the following parameters: $(m,n) = (2,-1)$, $(\Psi_T(v),\Psi_P(v)) = (1,0.5 + 0.25(v-0.5))$, $\beta = 0.05$, $\mathsf p(v)=1-v$, $\lambda = 10^{-3}$, and $(r_1^{\text{bottom}},r_1^{\text{top}}) = (-0.5,1.0)$. 
These parameters were chosen for a resonance at the surface $v(r_s) = 0.5$ with a rotational transform $\iota = 1/2$.
The leading-order solution is found by directly minimizing Eq.~\eqref{mean_potential_energy_simplified} on the undeformed domain, where $\theta = \zeta = 0$ and $v_0$ is represented by a degree-200 Legendre polynomial.
From the leading-order solution, we find the spatial location of the resonance is at $r_s \approx 0.493$ and the predicted layer width is $\lambda L_{mn}(r_s) \approx 4.33 \times 10^{-3}$.
The black line of Fig.~\ref{fig:asymptotic-solution} is a direct weak solution of the linearized energy principle \eqref{reduced_second_variation_slab}, also represented by a degree 200 Legendre polynomial. 
On top, we plot the asymptotic solution \eqref{uniform_soln_second_order} as a red dashed line, where the outer solution is obtained from Runge-Kutta integration of Eq.~\eqref{linearized_gse_fluct} with $\lambda=0$ and initial conditions at the rational surface.
We see excellent agreement between the two solutions.

\section{Nonlinear computation of statistical equilibria\label{sec:nonlinear_computation}}
We have just found that the statistical variational principle \eqref{mean_potential_energy_simplified} supports linearized solutions that smooth singular solutions by a length scale $\lambda$. 
The first purpose of this section is to show that this result extends to a fully nonlinear 3D problem. 
The second purpose is to show that the numerical solution to this problem converges exponentially to low tolerances, a property not satisfied by any existing 3D equilibrium solvers of the baseline model \eqref{MHS}. 

We begin by introducing the numerical method used to solve the equilibrium problem in Sec.~\ref{subsec:numerical-methodology}. 
Then, we introduce the 3D test problem in Sec.~\ref{subsec:3d-test-problem}, and finally we demonstrate the method in Sec.~\ref{subsec:results}

\subsection{Numerical Methodology}
\label{subsec:numerical-methodology}
The aim of the statistical variational principle is to learn a label mapping $G : Q \to Q_0$, where $G$ is varied over the space of diffeomorphisms between the two spaces.
The diffeomorphism requirement can be reduced to two requirements: (i) that $G$ is invertible and (ii) that $G$ maps the boundary of $Q$ to the boundary of $Q_0$. 
We have found that (i) is satisfied for all problems that we have tested, although we note that there is no guarantee that it will be satisfied for arbitrary boundary conditions. 
In contrast, the boundary condition requirement (ii) needs to be addressed directly. 

The standard method for enforcing boundary conditions in stellarator codes \cite{hirshman_steepest-descent_1983, dudt_desc_2020} is to compute a function $\lambda : Q_0 \to \mathbb R$ which defines an automorphism on the reference domain $(\rad,\theta,\zeta) \mapsto (\rad,\theta,\zeta+\tilde \lambda(\rad,\theta,\zeta))$.
This is composed with another mapping from $M : Q_0 \to Q$ to represent the whole function in real-space cylindrical coordinates.
In this way, one can apply Dirichlet boundary conditions to $M$ --- $M(\bm x) = M_0(\bm x)$ for $\bm x\in \partial Q_0$ --- while $\tilde \lambda$ is free to ``slide'' the magnetic coordinates along the boundary.
This process makes the solution nonlinear in the unknowns, which can be used to reduce the spectral width by a process called spectral condensation \cite{hirshman_optimized_1985}.
However, it also makes the solution a nonlinear function of the degrees of freedom, which adds complexity to the analysis.

Instead, we opt for a linear method of representing the solution on the slab.
Let $Q_c = [-1,1] \times \mathbb T^2$ be a computational domain.
We suppose that we have access to two known diffeomorphisms $\tilde G : Q_c \to Q$ and $\tilde G_0 : Q_c \to Q_0$ that map the computational domain onto the the spatial domain and reference domain.
In practice, we choose these maps to be 
\begin{align*}
    \tilde G(\rad_c, \theta_c, \zeta_c) &= \begin{pmatrix}
        \frac{1 + \rad_c}{2}r^{\mathrm{top}}(\theta_c, \zeta_c) + \frac{1 - \rad_c}{2}r^{\mathrm{bottom}}(\theta_c, \zeta_c)\\ \theta_c\\ \zeta_c
    \end{pmatrix}, 
    \\ \tilde G_0(\rad_c, \theta_c, \zeta_c) &= \begin{pmatrix}
        (\rad_c +1)/2 \\ \theta_c \\ \zeta_c
    \end{pmatrix},
\end{align*}
which linearly shift the radial coordinate to match the boundaries of $Q$ and $Q_0$ respectively.
Using these maps, we represent the solution as
\begin{equation*}
    G = \tilde G_0  \circ (\bm 1 + F) \circ \tilde G^{-1},
\end{equation*}
where $\bm 1 : Q_c \to Q_c$ is the identity map and $F : Q_c \to \mathbb R^3$ is the unknown component of $G$ that we will solve for.
When we pull the boundary condition back to the computational domain, it reduces to $F^{\rad_c}(\pm1,\theta_c, \zeta_c) = 0$, a linear Dirichlet constraint on the radial component of the mapping.
This is easily satisfied by restricting the basis of $F^{\rad_c}$.
We note that, while choosing $Q_c = Q_0$ would produce the same result, the domains are distinct in the general case.

We make this concrete by introducing a spectral discretization for $F : Q_0 \to \mathbb R^3$ by
\begin{equation*}
    F(\rad_c, \theta_c, \zeta_c) = \sum_{n_1=0}^{N_\rad-1} \sum_{n_2=0}^{N_\theta-1} \sum_{n_3 =0}^{N_\zeta-1} F_{n_1n_2n_3} \mathcal P_{n_1}(\rad_c) \mathcal F_{n_2}(\theta_c) \mathcal F_{n_3}(\zeta_c),
\end{equation*}
where $\mathcal P_n$ is the degree $n$ Legendre polynomial and $\mathcal F_n$ is the order $n$ Fourier mode, defined by
\begin{equation*}
    \mathcal F_n(\theta) = \begin{cases}
        \cos(n\theta/2), & n\text{ even}, \\
        \sin((n+1)\theta/2), & n\text{ odd}.
    \end{cases}
\end{equation*}
Clearly, the total number of degrees of freedom is $N = N_\rad N_\theta N_\zeta$.
Using the fact that $T_n(\pm 1) = (\pm 1)^n$, we represent the Dirichlet boundary conditions by eliminating $F_{0n_2n_3}$ and $F_{1n_2n_3}$ for all $0\leq n_2< N_\theta$ and $0\leq n_3 < N_\zeta$ via
\begin{equation*}
    F^{\rad_c}_{0n_2n_3} = -\sum_{m=1}^{\floor{\frac{N_\rad-1}{2}}} F^{\rad_c}_{2m,n_2,n_3}, \qquad F^{\rad_c}_{1n_2n_3} = - \sum_{m=1}^{\floor{N_\rad/2}-1} F^{\rad_c}_{2m+1,n_2,n_3}.
\end{equation*}
(Alternatively, other methods for eliminating the variables may provide better round off errors.)
A second constraint removes the symmetry in the choice of flux surface origin by requiring that the averaged quantities $\int F^{\theta_c} \df\theta_c \df\zeta_c$ and $\int F^{\zeta_c}\df \theta_c \df \zeta_c$ are both zero throughout the domain.
Using the orthogonality of Fourier series, these result in the two linear constraints that
\begin{equation*}
    F^{\theta_c}_{n_100} = F^{\zeta_c}_{n_100}= 0, \qquad \text{for all }0\leq n_1<N_\rad.
\end{equation*}
Let $\bm F$ be the vector of all of the reduced degrees of freedom, which has the total length $N = 3N_\rad N_\theta N_\zeta - 2 N_\theta N_\zeta - 2N_\rad$.

Returning to the variational problem, we phrase the discretized weak form of the statistical equilibrium problem as an unconstrained optimization problem in the unknowns
\begin{equation*}
    \min_{\bm F}\mathsf W[G],
\end{equation*}
where we use the nondimensional form of $\mathsf W$. 
To approximate the integrals over the phase space, we use Gauss-Legendre-Lobatto (GLL) quadrature in the radial direction and trapezoidal quadrature in the angular directions as
\begin{align*}
    \int_{Q}f(\bm x)\, \df^3 \bm x &= \int_{Q_c}f(G(\bm x_c)) \abs{ \dd{G}{\bm x_c} } \, \df^3 \bm x_c, \\
        &\approx \frac{(2\pi)^2}{M_2 M_3} \sum_{m_1=0}^{M_1-1} \sum_{m_2=0}^{M_2-1} \sum_{m_3=0}^{M_3-1}w_{m_1} \bigg[f(G(\bm x_c))\abs{\dd{G}{\bm x_c}}\bigg]_{\bm x_c=\bm x_{m_1m_2m_3}} ,
\end{align*}
where $\bm x_{m_1m_2m_3} = (\rad_{m_1},\theta_{m_2},\zeta_{m_3})$ are the 3D quadrature points, $(\rad_{m_1},w_{m_1})$ are the GLL quadrature points and weights in the radial coordinates, $\theta_{m_2} = 2\pi m_2/M_2$ are equispaced trapezoidal quadrature nodes over $\theta_c$, and $\zeta_{m_3}=2\pi m_3/M_3$ are the same over the $\zeta_c$ coordinate.
We always oversample the quadrature by $M_j = 2N_j+1$ for all $j$ to combat aliasing from nonlinearities in the energy principle.

To perform the optimization, we use the Limited-memory Broyden-Fletcher-Goldfarb-Shanno (L-BFGS) algorithm with Hager-Zhang line search provided by the \verb|Optim.jl| Julia optimization package.
The L-BFGS algorithm is a quasi-Newton method that minimizes $\mathsf W$ by approximating the Hessian $\dd{^2 \mathsf W}{\bm F^2}$ using only evaluations of the gradient $\dd{\mathsf W}{\bm F}$. 
The ``Limited-memory'' part of L-BFGS limits the storage to a specified number of gradient evaluations, which we set to be $100$ for most runs.
To accelerate the method, we use a block-Jacobi preconditioner, where the blocks consist of elements of the Hessian where the $\theta_c$ and $\zeta_c$ Fourier mode numbers are both equal. 
As we saw in the asymptotics, the Fourier mode numbers decouple in the undeformed configuration, and therefore the block-Jacobi Hessian is the exact Hessian for the undeformed configuration.
While this is not true far from equilibrium, it motivates our use of the preconditioner for this problem.
Because the preconditioner requires computing a significant part of the Hessian, we amortize the cost by only recomputing the preconditioner every $100$ iterations, aligning with the L-BFGS memory limit.

We consider the optimization converged when the norm of the Jacobian is lower than some tolerance $\norm{\dd{\mathsf W}{\bm F}} < \epsilon_{\mathrm{conv}}$.
It is important that the optimization is measured by this rather than, say, directly checking the value of $\mathsf W$. 
Near the minimum of $\mathsf W$, there are many $\mathcal O(10^{-8})$ high-frequency perturbations that have the same value of the objective to machine precision, but may significantly affect the magnetic current when a second derivative is applied.
We note that $\dd{\mathsf W}{\bm F}$ is the weak form of Force balance, and therefore our stopping criteria is a measure of how well force balance has been satisfied.
Unless otherwise stated, we choose $\epsilon_{\mathrm{conv}} = 10^{-10}$.

The code is written in pure Julia with GPU acceleration through the package \verb|CUDA.jl|.
All reported simulation timings were performed on a single NVIDIA RTX 6000 Ada Generation Graphics Card.

\subsection{A 3D Test Problem}
\label{subsec:3d-test-problem}

For the test problem, we choose a configuration with two low-order resonances in the rotational transform.
This is achieved by taking
\begin{equation*}
    \begin{pmatrix}
        \Psi_T(\rad) \\ \Psi_P(\rad)
    \end{pmatrix} = \begin{pmatrix}
        \cos(\gamma(\rad)) \\ \sin(\gamma(\rad))
    \end{pmatrix},
\end{equation*}
where a linear profile $\gamma(\rad) = 0.2532+0.1959q$ is chosen so that there are resonances in the rotational transform
\begin{equation*}
    \iota(\rad) = \frac{\Psi'_P}{\Psi'_T} = \tan(\gamma(\rad)),
\end{equation*}
of $\iota(0.35) = 1/3$ at $\iota(.65) = 2/5$. 
The rotational transform was chosen so that two rational resonances are obtained with a reasonably flat rotational transform profile.
In particular, we avoid the confounding effects of misses both the $\iota = 1/4$ and $\iota = 1/2$ resonances.
We choose a corresponding resonant top boundary condition of the form
\begin{equation*}
    r^{\mathrm{top}}(\theta,\zeta) = 1 + \epsilon(\cos(5\theta-2\zeta) + \cos(3\theta-\zeta)), \qquad r^{\mathrm{bottom}}(\theta,\zeta) = 0.
\end{equation*}
where once again $\epsilon$ is the magnitude of the boundary perturbation.
We fix the pressure of the equilibrium by setting $\beta = 0.05$ and $\mathsf p(r) = 1-r$.

The test problem has the following benefits 
\begin{enumerate}
    \item Genuinely three-dimensional. Because the boundary condition includes multiple Fourier modes, there is no symmetry direction that reduces the problem to a 2D Grad-Shafranov equation.
    \item Genuinely resonant. Because the boundary perturbations resonate with the rotational transform, we ensure that the problem clearly targets the differences between the MHD and statistical equilibrium models. 
    \item Genuinely nonlinear. By scanning the values of $\epsilon$, we can investigate the transition from perturbative to non-perturbative behavior of the solver, which appears as higher-order resonances in the Fourier spectrum.
\end{enumerate}

\subsection{Results}
\label{sec:phenomenology}
\label{subsec:results}
\begin{figure}
    \centering
    \includegraphics[width=0.9\textwidth]{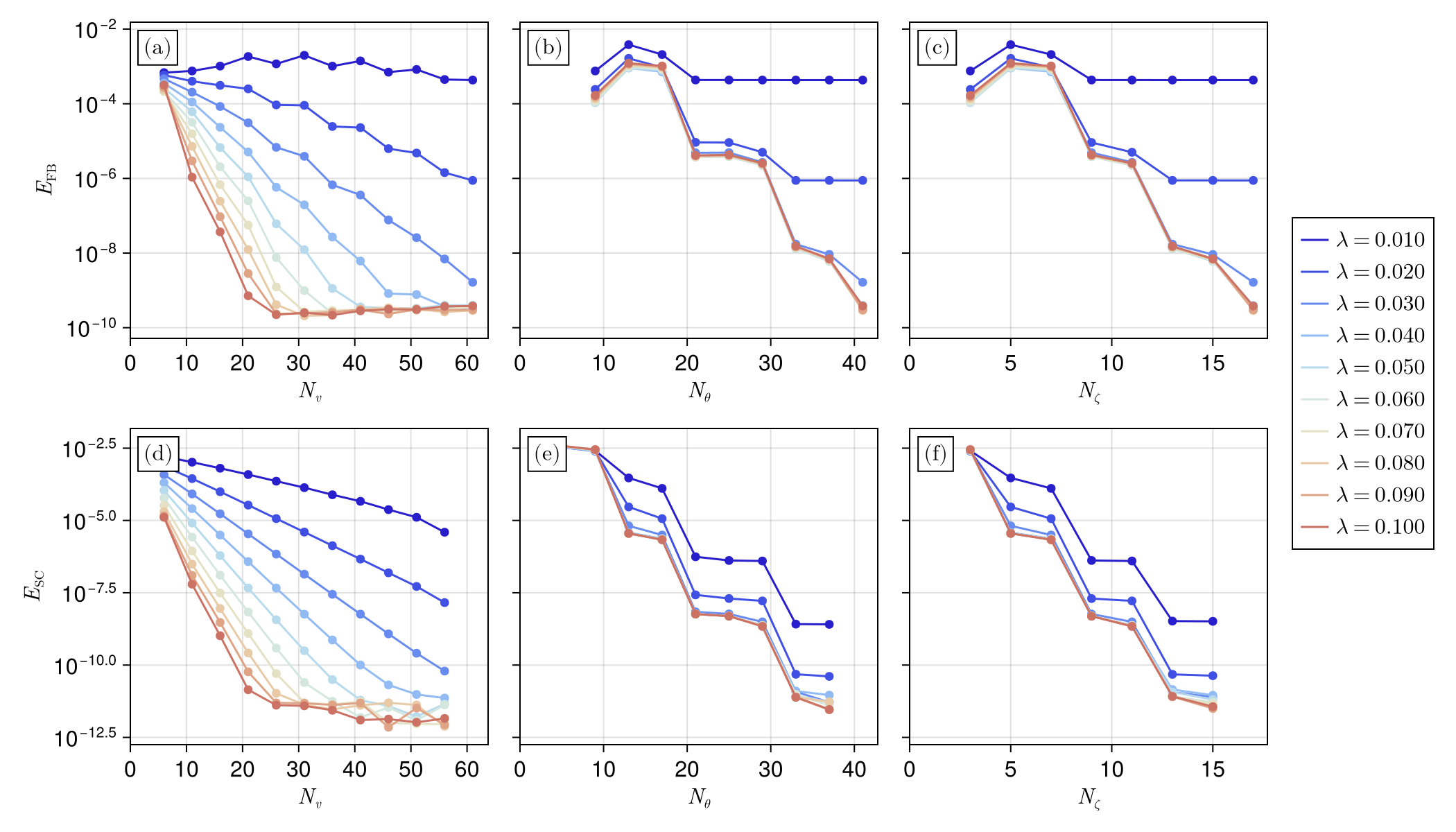}
    \caption{
    A plot demonstrating the convergence of the numerical method for $\epsilon=10^{-3}$.
    The top row shows convergence of the force-balance residual \eqref{eq:force-balance-residual} as a function of $\lambda$ and of the (a) radial, (b) toroidal, and (c) poloidal resolutions.
    The bottom row shows the self-convergence \eqref{eq:self-convergence-residual} of the solution of the variational equilibrium problem in the (d) radial, (e) toroidal, and (f) poloidal resolutions.
    }
    \label{fig:self-convergence}
\end{figure}
This section is organized into two categories: convergence and phenomenology.

\paragraph{Convergence} We begin by demonstrating the convergence of the numerical method on the test problem for an $\epsilon = 10^{-3}$ boundary perturbation.
To do so, for a fixed value of $\lambda > 0$, we consider the resolution of $(N_\rad^\star,N_\theta^\star,N_\zeta^\star) = (61,41,17)$ to be a high-fidelity simulation.
Here, the angular resolutions are chosen so that $(N_\zeta-1)/(N_\theta-1) = 2/5$, matching the larger rotational transform of the two resonances.
We set a solver tolerance of $\epsilon_{\mathrm{conv}}=10^{-12}$ to demonstrate low residuals.
We analyze convergence to the high-fidelity scale by individually scanning the three resolution parameters at lower values with the other two fixed.
For example, when testing convergence in the radial variable, we run simulations with $N_\rad < N_\rad^*$ and $(N_\theta,N_\zeta) = (N_\theta^\star,N_\zeta^\star)$.
We measure this convergence as $N_j \to N_j^\star$ in two ways.

The first measure is the $L^2$ norm of the force balance residual
\begin{equation}
\label{eq:force-balance-residual}
    E_{\mathrm{FB}} = \bigg[\frac{1}{(2\pi)^2}\int_Q \bigg| \nabla\cdot \bigg(p\mathbb{I}+\mu_0^{-1}\frac{1}{2}|\Bbar|^2\mathbb{I} - \mu_0^{-1}\Bbar\Bbar^T + \frac{1}{2}\mu_0^{-1}\text{tr}(\Sigma)\mathbb{I} - \mu_0^{-1}\Sigma\bigg)\bigg|^2 \,\df^3 \bm x\bigg]^{1/2}.
\end{equation}
This is a measure of how well we solve the statistical equilibrium PDE in the strong form.
We integrate the force balance residual over the quadrature grid associated with the high-resolution solution, meaning we evaluate force balance errors on at least eight times the number of quadrature points as there are degrees of freedom. 

While the first measure tells us whether we solve the PDE, it does not necessarily tell us whether the solution converges.
To address this, we additionally consider an $L^2$ measure of self-convergence.
Let $G^\star$ be the high-fidelity back-to-labels map, and let $G$ be a lower resolution map for the same problem.
Then, the self-convergence measure is
\begin{equation}
\label{eq:self-convergence-residual}
    E_{\mathrm{SC}} = \bigg[\frac{1}{(2\pi)^2} \int_Q |G(\bm x)-G^*(\bm x)|^2 \, \df^3 \bm x \bigg]^{1/2}.
\end{equation}
This integral is integrated over the quadrature grid of the high-resolution simulation.

In Fig.~\ref{fig:self-convergence}, we plot both measures of error as a function of both resolution and of the fluctuation parameter $\lambda$.
We see that the solutions self-converge to $\mathcal O(10^{-12})$ errors, which we note is the tolerance used to stop the L-BFGS optimization procedure. 
The error for the force balance residual is approximately three orders of magnitude worse, which we attribute to the multiple derivatives required to compute the magnetic field and current. 
In particular, we note that there is no convergence in the force balance residual for the lowest value of $\lambda$ considered, suggesting that it is possible to see weak self-convergence in the coordinates without converging to a strong solution. 

We see that the solutions with larger $\lambda$ converge at lower resolutions, indicating that highly fluctuating solutions are smoother.
Moreover, we remark that the lines of convergence appear linear in a log-linear plot.
This suggests that the numerical method is converging exponentially --- i.e.~the error goes as $\mathcal O(e^{-\alpha_\rad N_\rad - \alpha_\theta N_\theta - \alpha_\zeta N_\zeta})$ where $\alpha_\rad,\alpha_\theta,\alpha_\zeta > 0$. 
This is the type of convergence that is typically observed for spectral methods when the problem has smooth (analytic) solutions \cite{boyd_chebyshev_2001}.
Plots of the coefficient power spectrum agree with this point.
In sum, the convergence results strongly suggest that the statistical equilibrium problem is well-posed in 3D and that solutions are smooth. 
To our knowledge, there are no equivalent results for fully 3D instantiations of the MHD equilibrium problem.

\begin{figure}
    \centering
    \includegraphics[width=0.7\textwidth]{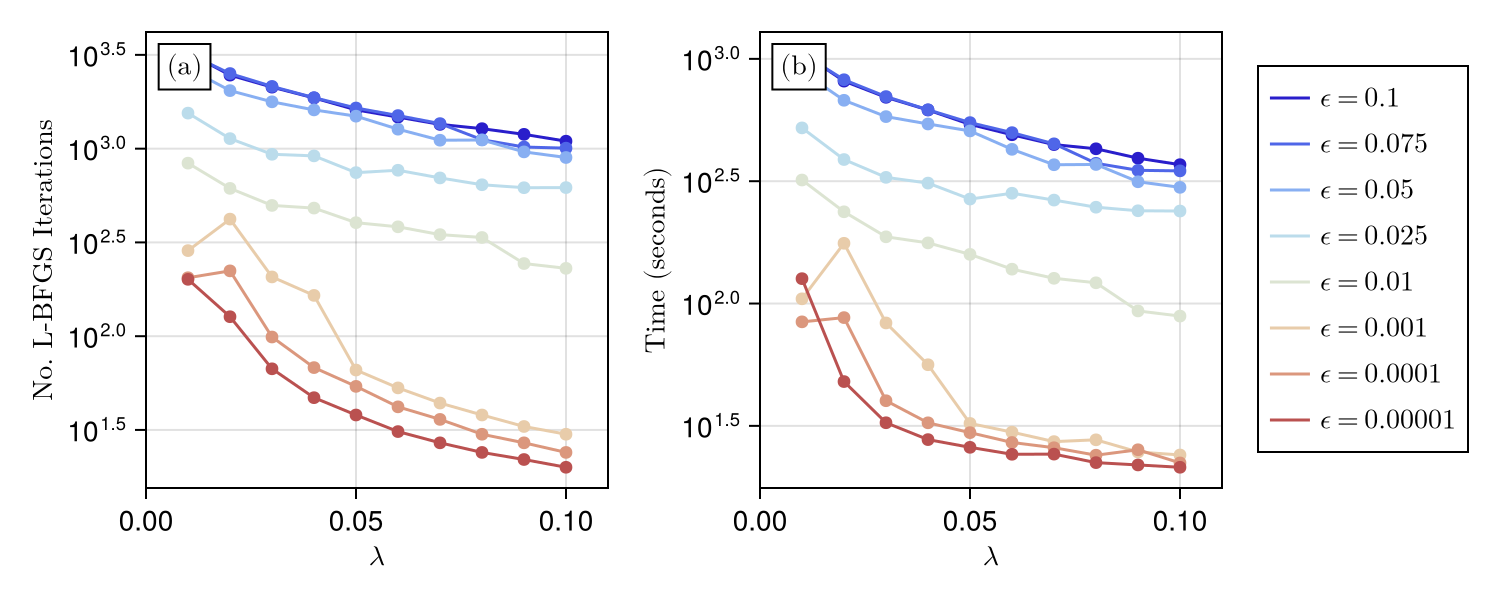}
    \caption{(a) Number of L-BFGS iterations to the solution and (b) the time to solution, both as a function of the fluctuation parameter $\lambda$ and perturbation magnitude $\epsilon$.
    }
    \label{fig:numerical-behavior}
\end{figure}

Beyond convergence under refinement, we also find that the L-BFGS optimization procedure is improved by the statistical equilibrium model (Fig.~\ref{fig:numerical-behavior}).
We test this by scanning over both fluctuation parameter $\lambda\in [0.01,0.1]$ on 10 uniform points and the boundary perturbation $\epsilon \in[10^{-6},10^{-1}]$.
The resolution parameters are $(N_r,N_\theta,N_\zeta) = (61,41,17)$.
For each case, we measure both the number of L-BFGS iterations required to reach the solution and the total time to reach the solution.
These values are plotted in panels (a) and (b) of Fig.~\ref{fig:numerical-behavior} respectively.
For all values of $\epsilon$, we find that increasing $\lambda$ leads to improved rates of convergence.
This is likely explained by improved conditioning of the preconditioned Hessian of the equilibrium problem as $\lambda$ increases. 

\begin{figure}
    \centering
    \includegraphics[width=0.9\textwidth]{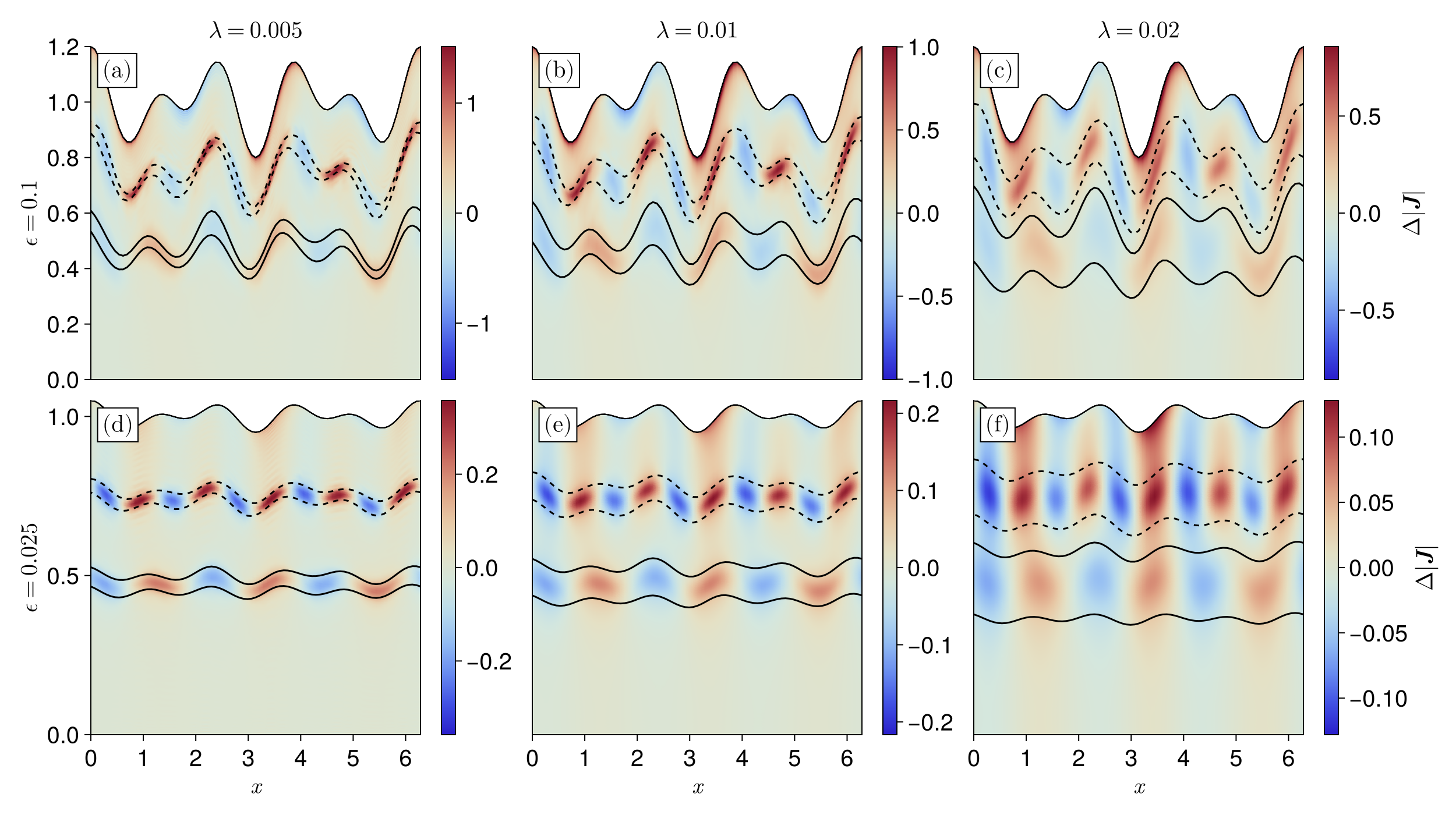}
    \caption{Current sheets plotted on the $y=0$ plane. The fluctuation parameter is scanned from left to right, with $\lambda =0.005$ (a,d), $\lambda = 0.01$ (b,e), and $\lambda=0.02$ (c,f). Two values of the perturbation magnitude, $\epsilon = 0.1$ (a,b,c) and $\epsilon=0.025$ (d,e,f), are plotted. Lines showing the predicted current sheet widths are plotted at $v=v_0(r_s) \pm \lambda v_0'(r_s) L_{mn}(r_s)$ for the resonant values of $v_0(r_s) \in \{0.35, 0.65\}$.}
    \label{fig:J-sheet}
\end{figure}

\paragraph{Phenomenology}
In Sec.~\ref{sec:asymptotics}, we observed from the linearized problem that the statistical equilibrium problem smooths singular solutions near rational surfaces.
Now, we address the question of whether this behavior persists to large deformations.

To clearly observe the current sheets, we first need to subtract off the background current from the simulations.
Consider a back-to-labels map $G_{\lambda,\epsilon}$, which is the solution to the PDE with fluctuation parameter $\lambda$ and boundary perturbation magnitude $\epsilon$. 
A clean representation of current sheets for this map is found by measuring the deviation between the current predicted from this map $\bm J_{\lambda, \epsilon}(\bm x)$ and the current from the undeformed configuration $\bm J_{\lambda, 0}(\bm x)$. 
We reduce this to a single field by subtracting the norms in the reference domain $\Delta|\bm J| = |\bm J_{\lambda,\epsilon} \circ G_{\lambda,\epsilon}| - |\bm J_{\lambda,0} \circ G_{\lambda_0}|$.
The result is that the mean, undeformed current is removed from the solution, leaving only the deviation.

In Fig.~\ref{fig:J-sheet}, we plot the current deviation on the $y = 0$ surface in the spatial domain for three values of the fluctuation parameter $\lambda \in \{0.005, 0.01, 0.02\}$ and two values of the perturbation magnitude $\epsilon \in \{0.025,0.1\}$. 
The simulations are all run with a resolution $(N_\rad, N_\theta, N_\zeta) = (101, 41, 17)$.
On top of the current deviation, we plot contours flux label $v = v_0(r_s) \pm \lambda v_0'(r_s)L_{mn}(r_s)$, where $v_0$ is the undeformed solution, $v_0(r_s) \in \{0.35,0.65\}$ is the location of the resonant rotational transform, and $\lambda v_0'(r_s)L_{mn}(r_s)$ is the predicted asymptotic length scale of the smoothed singularity in Eq.~\eqref{eq:layer-width} in the reference domain.
Visually, we find that the nonlinear 3D current sheets scale with $\lambda$ according to the length scale predicted by the asymptotics.
This is true even in the $\epsilon = 0.1$ case, which is far from perturbative.

\section{Improved energy landscape\label{sec:improved_energy_landscape}}
For axisymmetric magnetic fields the usual MHD equilibrium model reduces to a single scalar PDE for a flux function known as the Grad-Shafranov equation. This reduction---from a PDE system to a single scalar equation---dramatically simplifies mathematical analysis of the axisymmetric equilibrium model. Weitzner \cite{weitzner_ideal_2014} recognized that reduction to a single scalar equation also works in $3D$, but advocated against it because he did not believe doing so leads to a viable iterative solution method for the overall MHD equilibrium model. Grad-Shafranov reduction also applies to our new statistical equilibrium model in $3D$. In spite of Weitzner's words of caution, the procedure he identified for constructing a $3D$ Grad-Shafranov equation, entailing the elimination of the angular components of $G$, is mathematically sound, even if not useful for iterative solution methods in the MHD equilibrium model. We may therefore compare the analytical properties of the MHD equilibrium model ($\lambda= 0$) with those of the statistical equilibrium model ($\lambda > 0$) by comparing their respective Grad-Shafranov equations. This section applies this idea to study the second variation of the potential energy at small scales in both MHD and statistical equilibrium. The analysis will reveal the non-ellipticity of the MHD equilibrium model and the ellipticity of the statistical equilibrium model. This implies the small scale smoothing observed earlier, both analytically and numerically, is a general feature of the model.
\subsection{MHD equilibrium}
First consider the usual MHD equilibrium model, characterized by $\lambda = 0$. Assemble the angle variables into the vector $\vartheta= (\theta,\zeta)^T$. The potential energy can be written as
\begin{equation*}
    W(\rad,\vartheta) = \frac{1}{2}\mathcal{Q}(\rad)[\vartheta,\vartheta] - \beta\,\int \mathsf{p}(\rad)\,d^3\bm{x},
\end{equation*}
where the $\rad$-dependent quadratic form $\mathcal{Q}(\rad)$ is given by
\begin{align*}
    \mathcal{Q}(\rad)[\vartheta_1,\vartheta_2] & =\int \bigg(\begin{pmatrix}\nabla & 0\\ 0 & \nabla \end{pmatrix}\vartheta_1\bigg)^T M\begin{pmatrix} \nabla \rad_{\times}^T\nabla\rad_\times & 0 \\ 0 &\nabla \rad_{\times}^T\nabla\rad_\times \end{pmatrix} \begin{pmatrix}\nabla & 0\\ 0 & \nabla \end{pmatrix}\vartheta_2\,d^3\bm{x}.
\end{align*}
Here we have introduced the positive semi-definite $6\times 6$ matrix 
\begin{align}
    M = \begin{pmatrix} (\Psi_T')^2 \mathbb I_3 & -\Psi_T'\Psi_P' \mathbb I_3\\ -\Psi_T'\Psi_P' \mathbb I_3 & (\Psi_P')^2 \mathbb I_3\end{pmatrix},\label{dynamical_M}
\end{align}
where $\mathbb I_3$ is the $3\times 3$ identity matrix.
In weak form, the $\vartheta$-Euler-Lagrange equation is 
\begin{align}
    \mathcal{Q}(\rad)[\delta\vartheta,\vartheta] = 0,\quad \forall\,\delta\vartheta.\label{theta_ele_weak}
\end{align}  
In strong form this condition is equivalent to the PDE system
\begin{align}
    \begin{pmatrix}\nabla\cdot & 0\\ 0 & \nabla\cdot \end{pmatrix}\bigg(M\begin{pmatrix} \nabla \rad_{\times}^T\nabla\rad_\times & 0 \\ 0 &\nabla \rad_{\times}^T\nabla\rad_\times \end{pmatrix} \begin{pmatrix}\nabla & 0\\ 0 & \nabla \end{pmatrix}\vartheta\bigg) = 0.\label{theta_ele_strong}
\end{align}
Using the vector identity $\nabla\rad_\times^T \nabla\rad_\times = |\nabla \rad|^2 \mathbb I_3 -\nabla \rad \nabla \rad^T$, the last equation may also be written as the system
\begin{align}
0=&\nabla_s\cdot\bigg((\Psi_T'\Psi_T')|\nabla\rad|\nabla_s\theta-\Psi_T'\Psi_P'|\nabla \rad|\nabla_s \zeta\bigg),\label{dyn_ang_pde_one}\\
0 =& \nabla_s\cdot\bigg(-\Psi_P'\Psi_T'|\nabla\rad|\nabla_s\theta+(\Psi_P'\Psi_P')|\nabla \rad|\nabla_s \zeta\bigg),\label{dyn_ang_pde_two}
\end{align}
where $\nabla_s$ and $(\nabla_s\cdot{})$ denote the $\rad$-surface gradient and divergence, defined by 
\begin{align*}
    \nabla_s=\bigg(\mathbb I_3 -\frac{\nabla \rad\nabla \rad^T}{|\nabla \rad|^2}\bigg)\nabla, \qquad 
    \nabla_s\cdot\star = |\nabla \rad|^{-1} \nabla\cdot\bigg[\bigg(\mathbb I_3 -\frac{\nabla \rad\nabla \rad^T}{|\nabla \rad|^2}\bigg) |\nabla \rad| \star\bigg].
\end{align*}
Since $\Psi_T'$ and $\Psi_P'$ are pure functions of $\rad$, they commute with both the surface gradient and divergence, making Eq.~\eqref{dyn_ang_pde_one} a multiple of Eq.~\eqref{dyn_ang_pde_two}. In other words, Eqs.~\eqref{dyn_ang_pde_one}-\eqref{dyn_ang_pde_two} comprise a single PDE instead of a system. This justifies introducing the additional equation
\begin{align}
    \nabla_s\cdot(|\nabla\rad|\nabla_s\zeta) = 0,\label{break_degeneracy_zeta}
\end{align}
to break the degeneracy, implying that $\theta$ must satisfy the same equation,
\begin{align}
\nabla_s\cdot(|\nabla\rad|\nabla_s\theta) = 0.\label{break_degeneracy_theta}
\end{align}
The system \eqref{break_degeneracy_zeta}-\eqref{break_degeneracy_theta} is really a family of $2D$ elliptic PDEs parameterized by flux surface label $\rad$. After specifying the periods,
\begin{align}
&\oint_{T}\nabla_s\theta\cdot d\ell = 0,\quad \oint_P\nabla_s\theta\cdot d\ell = 2\pi,\label{periods}\\
&\oint_{T}\nabla_s\zeta\cdot d\ell = 2\pi,\quad \oint_P\nabla_s\zeta\cdot d\ell = 0,\nonumber
\end{align}
and specifying a conventional $\vartheta=0$ point on each surface, it has a unique $\rad$-dependent solution, $\vartheta = \Theta(\rad) = (\hat{\theta}(\rad),\hat{\zeta}(\rad))^T$. 

Eliminate $\vartheta$ within the variational principle by introducing the Grad-Shafranov functional $w(\rad) = W(\rad,\Theta(\rad))$. Note that the chain rule and the definition of $\Theta(\rad)$ together imply the useful identity
\begin{align}
    \delta w(\rad)[\delta\rad] = \frac{1}{2}(\delta\mathcal{Q}(\rad)[\delta \rad])[\Theta(\rad),\Theta(\rad)] - \beta\int \mathsf{p}'(\rad)\,\delta \rad\,d^3\bm{x}.\label{dyn_abs_identity}
\end{align}
The weak form of the Euler-Lagrange equation for $w$ is therefore
\begin{align*}
    0& = \frac{1}{2}(\delta\mathcal{Q}(\rad)[\delta \rad])[\Theta(\rad),\Theta(\rad)] - \beta\int \mathsf{p}'(\rad)\,\delta \rad\,d^3\bm{x} ,\quad \forall\delta\rad.
\end{align*} 
A direct calculation leads to the strong form of this equation,
\begin{gather}
    \nabla\cdot\bigg(\mu\nabla \rad\bigg) + \beta\mathsf{p}'(\rad) = (\Psi_T''\nabla\hat{\theta} - \Psi_P''\nabla\hat{\zeta})^T(\nabla\rad^T_\times\nabla\rad_\times)(\Psi_T'\nabla\hat{\theta} - \Psi_P'\nabla\hat{\zeta}),\label{dyn_gse}
\end{gather}
where we have introduced the positive semi-definite $3\times 3$ matrix
\begin{align*}
    \mu = (\Psi_T'\nabla\hat{\theta} - \Psi_P'\nabla\hat{\zeta})_\times^T(\Psi_T'\nabla\hat{\theta} - \Psi_P'\nabla\hat{\zeta})_\times .
\end{align*}
We will refer to Eq.\,\eqref{dyn_gse} as the Grad-Shafranov equation because it is a scalar equation for the flux function $\rad$ that is equivalent to the full MHD equilibrium model, just like the usual Grad-Shafranov equation in the axisymmetric setting. 
It also coincides with the original $\rad$-Euler-Lagrange equations for $W$ with $\vartheta = \Theta(\rad)$, thereby confirming validity of the variational principle defined by ${w}$. In other words, we can dispense with the three-field variational principle defined by $W$; all solutions of the MHD equilibrium model can be obtained as critical points of the one-field Grad-Shafranov variational principle $\delta w(\rad) = 0$.

The most tractable variational principles in mathematical physics enjoy bowl-shaped energy landscapes near critical points, meaning positive-definiteness of the second variation. If the Grad-Shafranov functional is bowl-shaped near a solution $\rad$, then the second variation $\delta^2 w(\rad)[\delta\rad^*,\delta\rad]$ must therefore be positive definite. This type of stability analysis generally requires numerical methods to make progress. However, the WKB ansatz $\delta \rad = \delta \overline{\rad}\,\exp(iS)$, where $S$ is a rapidly-varying phase, makes the sign of $\delta^2 w$ tractable at small length scales. This method also assesses ellipticity of \eqref{dyn_gse} as a convenient byproduct.

Before proceeding with the short-scale analysis, compute an exact expression for the second variation of the Grad-Shafranov functional as follows. First vary \eqref{dyn_abs_identity} in $\rad$ along $\delta \rad^*$ to obtain
\begin{align*}
    \delta^2 w(\rad)[\delta\rad^*,\delta\rad] &  = \frac{1}{2}(\delta^2\mathcal{Q}(\rad)[\delta\rad^*,\delta\rad])[\Theta(\rad),\Theta(\rad)] + (\delta\mathcal{Q}(\rad)[\delta\rad])[\Theta(\rad),\delta\Theta(\rad)[\delta\rad^*]]\\
    &- \beta\int \mathsf{p}''(\rad)\,|\delta\rad|^2\,d^3\bm{x}.
\end{align*}
Next notice that \eqref{theta_ele_weak} implies $\mathcal{Q}(\rad)[\delta\vartheta,\Theta(\rad)]=0$ for all $\rad$ and $\delta\vartheta$. Differentiate this result implicity to find $(\delta\mathcal{Q}(\rad)[\delta\rad])[\delta\vartheta,\Theta(\rad)] + \mathcal{Q}(\rad)[\delta\vartheta,\delta\Theta(\rad)[\delta\rad]]=0$ for all $\rad$ and $\delta\vartheta$. Then set $\delta\vartheta = \delta\Theta(\rad)[\delta\rad^*]$ to reveal the identity
\begin{align*}
    (\delta\mathcal{Q}(\rad)[\delta\rad])[\Theta(\rad),\delta\Theta(\rad)[\delta\rad^*]] = - \mathcal{Q}(\rad)[\delta\Theta(\rad)[\delta\rad^*],\delta\Theta(\rad)[\delta\rad]].
\end{align*}
Substitute back into the earlier formula for $\delta^2 w$ to finally conclude
\begin{align}
    \delta^2 w(\rad)[\delta\rad^*,\delta\rad] & =\frac{1}{2}(\delta^2\mathcal{Q}(\rad)[\delta\rad^*,\delta\rad])[\Theta(\rad),\Theta(\rad)] - \mathcal{Q}(\rad)[\delta\Theta(\rad)[\delta\rad^*],\delta\Theta(\rad)[\delta\rad]] \nonumber\\
    &- \beta\int \mathsf{p}''(\rad)\,|\delta\rad|^2\,d^3\bm{x},\label{second_var_general}
\end{align}
which is manifestly symmetric in $\delta\rad,\delta\rad^*$.

To find an explicit expression for the short-scale behavior of the second variation, begin by introducing the eikonal ansatz for $\delta\rad = \delta\overline{\rad}\,\exp(iS)$. Perform all calculations to leading-order in $\bm{k} = \nabla S$. The eikonal form for $\delta\rad$ implies $\delta\Theta(\rad)[\delta\rad] = \delta\overline{\Theta}\,\exp(iS)$ also has eikonal form. An explicit expression for $\delta\overline{\Theta}$ follows from implicit differentiation of $\mathcal{Q}(\rad)[\delta\vartheta,\Theta(\rad)] =0 $ and application of the eikonal ansatz. The result is
\begin{align}
    \delta\overline{\Theta} & = \frac{\delta\overline{\rad}}{|\nabla\rad|^2|\bm{k}_s|^2}\begin{pmatrix} (\bm{k}_\times^T\bm{k}_\times\nabla\rad)^T & 0 \\ 0 &(\bm{k}_\times^T\bm{k}_\times\nabla\rad)^T \end{pmatrix} \begin{pmatrix}\nabla & 0\\ 0 & \nabla \end{pmatrix}\Theta(\rad),\label{eikonal_delta_theta}
\end{align}
where $\bm{k}_s = \bm{k} - \bm{k}\cdot\nabla\rad\,\nabla\rad/|\nabla\rad|^2$ denotes the component of $\bm{k}$ tangential to the $\rad$-surface. Next note the high-$\bm{k}$ identities
\begin{align*}
    &\frac{1}{2}(\delta^2\mathcal{Q}(\rad)[\delta\rad^*,\delta\rad])[\Theta(\rad),\Theta(\rad)] \\
    & = \int |\delta\overline{\rad}|^2\bigg(\begin{pmatrix}\nabla & 0\\ 0 & \nabla \end{pmatrix}\Theta\bigg)^T M\begin{pmatrix} \bm{k}_{\times}^T\bm{k}_\times & 0 \\ 0 &\bm{k}_{\times}^T\bm{k}_\times \end{pmatrix} \begin{pmatrix}\nabla & 0\\ 0 & \nabla \end{pmatrix}\Theta\,d^3\bm{x}
\end{align*}
\begin{align*}
    &\mathcal{Q}(\rad)[\delta\Theta(\rad)[\delta\rad^*],\delta\Theta(\rad)[\delta\rad]]\\
    & = \int \bigg(\begin{pmatrix}\bm{k} & 0\\ 0 & \bm{k} \end{pmatrix}\delta\overline{\Theta}^*\bigg)^T M\begin{pmatrix} \nabla \rad_{\times}^T\nabla\rad_\times & 0 \\ 0 &\nabla \rad_{\times}^T\nabla\rad_\times \end{pmatrix} \begin{pmatrix}\bm{k} & 0\\ 0 & \bm{k} \end{pmatrix}\delta\overline{\Theta}\,d^3\bm{x}\\
    & = \int |\delta\overline{\rad}|^2\,\bigg(\begin{pmatrix} \nabla & 0 \\ 0 & \nabla\end{pmatrix}\Theta(\rad)\bigg)^TM\begin{pmatrix} \bm{k}_\times^T\bm{e}\bm{e}^T\bm{k}_\times& 0\\ 0 & \bm{k}_\times^T\bm{e}\bm{e}^T\bm{k}_\times \end{pmatrix} \begin{pmatrix}\nabla & 0\\ 0 & \nabla \end{pmatrix}\Theta(\rad)\,d^3\bm{x},
\end{align*}
where we have introduced the unit vector tangent to the $\rad$-surfaces
\begin{align*}
    \bm{e} = \frac{\bm{k}_s\times\nabla\rad}{|\bm{k}_s||\nabla\rad|}.
\end{align*}
Finally, use the general formula \eqref{second_var_general} for the second variation and the expression \eqref{eikonal_delta_theta} to find
\begin{align}
    \delta^2 w(\rad)[\delta\rad^*,\delta\rad] 
    & = \int |\delta\overline{\rad}|^2\sigma(\bm{x},\bm{k})\,d^3\bm{x},\nonumber
\end{align}
where we have identified the principal symbol \cite{hormanderAnalysisLinearPartial2007,taylorPartialDifferentialEquations2023} of the Grad-Shafranov equation:
\begin{align}
    \sigma(\bm{x},\bm{k}) &= |\bm{k}|^2\begin{pmatrix}\nabla\hat{\theta}\\ \nabla\hat{\zeta}\end{pmatrix}^T M  \begin{pmatrix}
        \bm e \bm e^T & 0 \\ 0 & \bm e \bm e^T
    \end{pmatrix}\begin{pmatrix}\nabla\hat{\theta}\\ \nabla\hat{\zeta}\end{pmatrix}\nonumber\\
    &= |\bm{k}|^2 \frac{(\bm{k}\cdot\bm{B})^2 }{|\bm{k}\times\nabla\rad|^2}.
\end{align}
Notice that when $\bm{k}_s = 0$ the principal symbol is ill-defined due to ambiguity in the direction of $\bm{e}$.

This calculation reveals a complex energy landscape for the Grad-Shafranov equation at small scales. The short-scale second variation may be written
\begin{align*}
    \delta^2 w(\rad)[\delta\rad^*,\delta\rad] = \int |\delta\overline{\rad}|^2 |\bm{k}|^2 \frac{(\bm{k}\cdot\bm{B})^2 }{|\bm{k}\times\nabla\rad|^2}\,d^3\bm{x}.
\end{align*}
Suppose there is some rational flux surface, where $\Psi_P'/\Psi_T' = -n/m$ for integers $n,m$. Choose a phase function of the form $S(\bm{x}) =s(m\theta + n\zeta)$. Note that $\bm{k}\cdot\bm{B}$ then vanishes along the rational surface. Also choose $\delta \rad$ to ensure $|\delta\overline{\rad}|^2$ approximates a $\delta$-function near that flux surface. Then $\delta^2w(\rad)[\delta\rad^*,\delta\rad]\approx 0$ becomes arbitrarily close to zero even though the norm of $\delta\rad$ is not small. This implies the presence of ``flat spots" in the energy landscape for the MHD equilibrium model -- directions along which one can perturb a solution without incurring an energy penalty. These flat spots may be understood as the leading cause of the well-known mathematical problems that caused H. Grad to formulate his famous conjecture \cite{grad_toroidal_1967} ruling out $3D$ MHD equilibria with nested flux surfaces.

\subsection{Statistical equilibrium}
Now consider the statistical equilibrium model with $\lambda > 0$. The analysis parallels that of the $\lambda = 0$ case, but with important differences. The averaged potential energy is $\overline{W}(\rad,\vartheta) = \frac{1}{2}\mathcal{Q}(\rad)[\vartheta,\vartheta] - \beta\int \mathsf{p}(\rad)\,d^3\bm{x}$, where the $\rad$-dependent quadratic form $\mathcal{Q}(\rad)$ is defined as in the $\lambda=0$ case but with 
\begin{align*}
    M = \begin{pmatrix} ((\Psi_T')^2+\lambda^2) \mathbb I_3& -\Psi_T'\Psi_P' \mathbb I_3\\ -\Psi_T'\Psi_P' \mathbb I_3 & ((\Psi_P')^2+\lambda^2) \mathbb I_3\end{pmatrix}.
\end{align*}
In contrast to the $\lambda =0$ case, this $M$ is positive definite for all $\lambda > 0$. The $\vartheta$-Euler-Lagrange equation is formally identical to \eqref{theta_ele_strong}. In components we find
\begin{align}
0=&\nabla_s\cdot\bigg((\Psi_T'\Psi_T'+\lambda^2)|\nabla\rad|\nabla_s\theta-\Psi_T'\Psi_P'|\nabla \rad|\nabla_s \zeta\bigg)\label{ang_pde_one}\\
0 =& \nabla_s\cdot\bigg(-\Psi_P'\Psi_T'|\nabla\rad|\nabla_s\theta+(\Psi_P'\Psi_P'+\lambda^2)|\nabla \rad|\nabla_s \zeta\bigg).\label{ang_pde_two}
\end{align}
Positive definiteness of $M$ and the lack of any radial derivatives of $\vartheta$ implies that this is really a family of $2D$ elliptic PDEs parameterized by flux surface label $\rad$ equivalent to Eqs.\,\eqref{break_degeneracy_zeta}-\eqref{break_degeneracy_theta}. After specifying the periods as in 
Eq.\,\eqref{periods},
and specifying a conventional $\vartheta=0$ point on each surface, it has a unique $\rad$-dependent solution, $\vartheta = \Theta(\rad) = (\hat{\theta}(\rad),\hat{\zeta}(\rad))^T$. 

Eliminate $\vartheta$ within the variational principle by introducing the statistical Grad-Shafranov functional $\overline{w}(\rad) = \overline{W}(\rad,\Theta(\rad))$. The strong form of the Euler-Lagrange equation is \eqref{dyn_gse}, but
where the $3\times 3$ matrix $\mu$ is now replaced with the positive definite matrix given by
\begin{align}
    \overline{\mu} = (\Psi_T'\nabla\hat{\theta} - \Psi_P'\nabla\hat{\zeta})_\times^T(\Psi_T'\nabla\hat{\theta} - \Psi_P'\nabla\hat{\zeta})_\times + \lambda^2 \nabla\hat{\theta}_\times^T\nabla\hat{\theta}_\times + \lambda^2\nabla\hat{\zeta}_\times^T\nabla\hat{\zeta}_\times.
\end{align}
We will refer to Eq.\,\eqref{dyn_gse} with this modified $\mu$ as the statistical Grad-Shafranov equation. As before, we can dispense with the three-field variational principle defined by $\overline{W}$ in favor of the statistical Grad-Shafranov variational principle $\delta\overline{w}(\rad) = 0$.


The second variation of $\overline{w}$ is still given by the formula \eqref{second_var_general}. We may therefore proceed directly to the short-scale stability analsyis. As before, begin by introducing the eikonal ansatz for $\delta\rad = \delta\overline{\rad}\,\exp(iS)$. Perform all calculations to leading-order in $\bm{k} = \nabla S$. The eikonal form for $\delta\rad$ implies $\delta\Theta(\rad)[\delta\rad] = \delta\overline{\Theta}\,\exp(iS)$ also has eikonal form. An explicit expression for $\delta\overline{\Theta}$ follows from implicit differentiation of $\mathcal{Q}(\rad)[\delta\vartheta,\Theta(\rad)] =0 $ and applying the eikonal ansatz. The result is
\begin{align*}
    M\begin{pmatrix} \bm{k}^T\nabla \rad_{\times}^T\nabla\rad_\times\bm{k} & 0 \\ 0 & \bm{k}^T\nabla\rad_{\times}^T\nabla\rad_\times\bm{k} \end{pmatrix} \delta\overline{\Theta}
    + \delta\overline{\rad}\,M\begin{pmatrix} \bm{k}^T\nabla \rad_{\times}^T\bm{k}_\times & 0 \\ 0 &\bm{k}^T\nabla \rad_{\times}^T\bm{k}_\times \end{pmatrix} \begin{pmatrix}\nabla & 0\\ 0 & \nabla \end{pmatrix}\Theta(\rad) = 0.
\end{align*}
Multiplying through by $M^{-1}$ (justified by positive-definiteness) and unpacking cross products leads to (somewhat surprisingly) Eq.\,\eqref{eikonal_delta_theta}.
Finally, again proceeding much as before, we use the general formula \eqref{second_var_general} for the second variation and the expression \eqref{eikonal_delta_theta} to find
\begin{align}
    \delta^2\overline{w}(\rad)[\delta\rad^*,\delta\rad] 
    & = \int |\delta\overline{\rad}|^2\overline{\sigma}(\bm{x},\bm{k})\,d^3\bm{x},\label{high_k_d2w}
\end{align}
where we have identified the principal symbol of the statistical Grad-Shafranov equation:
\begin{align}
    \overline{\sigma}(\bm{x},\bm{k}) &= |\bm{k}|^2
    \begin{pmatrix}\nabla\hat{\theta}\\ \nabla\hat{\zeta}\end{pmatrix}^T
    M 
    \begin{pmatrix} \bm e^T \bm e & 0 \\ 0 & \bm e \bm e^T \end{pmatrix}
    \begin{pmatrix}\nabla\hat{\theta}\\ \nabla\hat{\zeta}\end{pmatrix}\nonumber\\
    &= |\bm{k}|^2 \frac{(\bm{k}\cdot\bm{B})^2 + \lambda^2(\bm{k}\times\nabla\rad\cdot\nabla\hat{\theta})^2 + \lambda^2(\bm{k}\times\nabla\rad\cdot\nabla\hat{\zeta})^2}{|\bm{k}\times\nabla\rad|^2}.
\end{align}

This calculation reveals a bowl-shaped energy landscape for the statistical Grad-Shafranov equation at small scales. To see this let $\bm{\alpha} = \bm{k}\times \nabla\rad$ and denote its contravariant components $\bm{\alpha} = \alpha^\theta\,\partial_\theta + \alpha^\zeta\,\partial_\zeta$. (It has no $\partial_\rad$-component by tangency to the $\rad$-surfaces.) The squared length of the vector $\alpha$ is
\begin{align*}
    |\bm{\alpha}|^2 = \begin{pmatrix} \alpha^\theta \\ \alpha^\zeta\end{pmatrix}^Th\begin{pmatrix} \alpha^\theta \\ \alpha^\zeta\end{pmatrix},\quad h = \begin{pmatrix} \partial_\theta\cdot\partial_\theta & \partial_\theta\cdot\partial_\zeta\\ \partial_\theta\cdot\partial_\zeta & \partial_\zeta\cdot\partial_\zeta\end{pmatrix}.
\end{align*}
We can bound this from above using the Cauchy-Schwarz inequality
\begin{align*}
    |\bm{\alpha}|^2 = \text{tr}\left(\begin{pmatrix} \alpha^\theta \\ \alpha^\zeta\end{pmatrix}\begin{pmatrix} \alpha^\theta \\ \alpha^\zeta\end{pmatrix}^Th\right)\leq [(\alpha^\theta)^2 + (\alpha^\zeta)^2]|h|,
\end{align*}
where $|h| = \sqrt{\text{tr}(h^2)}$ denotes the Frobenius norm of $h$. The principal symbol $\overline{\sigma}(\bm{x},\bm{k})$ therefore admits the lower bound,
\begin{align}
\nonumber
    \overline{\sigma}(\bm{x},\bm{k}) & = |\bm{k}|^2 \frac{(\bm{k}\cdot\bm{B})^2 + \lambda^2(\bm{k}\times\nabla\rad\cdot\nabla\hat{\theta})^2 + \lambda^2(\bm{k}\times\nabla\rad\cdot\nabla\hat{\zeta})^2}{|\bm{k}\times\nabla\rad|^2}\\
\nonumber
    & \geq \lambda^2|\bm{k}|^2 \frac{ (\bm{k}\times\nabla\rad\cdot\nabla\hat{\theta})^2 + (\bm{k}\times\nabla\rad\cdot\nabla\hat{\zeta})^2}{|\bm{k}\times\nabla\rad|^2}\\
\label{eq:statistical-ellipticity}
    &\geq\lambda^2|\bm{k}|^2\frac{(\alpha^\theta)^2 + (\alpha^\zeta)^2}{[(\alpha^\theta)^2 + (\alpha^\zeta)^2]|h|} = \frac{\lambda^2}{|h|}|\bm{k}|^2,
\end{align}
which implies ellipticity of the statistical Grad-Shafranov equation.
Note that if $|h|$ is uniformly bounded from above then the previous inequality implies that the symbol $\overline{\sigma}(\bm{x},\bm{k})$ is elliptic.
This bound is satisfied under the condition that $G$ is a diffeomorphism.
Substituting this inequality into the short-scale second variation formula \eqref{high_k_d2w} leads to
\begin{align*}
    \delta^2\overline{w}(\rad)[\delta\rad^*,\delta\rad] \geq \int |\delta\overline{\rad}|^2\frac{\lambda^2}{|h|}|\bm{k}|^2\,d^3\bm{x} ,\quad \delta\rad = \delta\overline{\rad}\exp(iS),\quad \bm{k} =\nabla S.
\end{align*}
Thus, whenever the amplitude $\delta\overline{\rad}$ and the wave vector $\bm{k}$ are each non-zero the second variation is positive at short scales. Apparently replacing the MHD equilibrium model with the statistical equilibrium model eliminates all flat spots in the energy landscape associated with rational flux surfaces. This provides a compelling theoretical explanation for the favorable numerical and asymptotic properties of the statistical equilibrium model discussed in this Article.

\section{Discussion}
\label{sec:discussion}
We have demonstrated that the assumption of a quickly fluctuating nonideal magnetic field leads to a statistical equilibrium principle \eqref{mean_potential_energy}.
The particular scenario of an ideal gas with vanishing adiabatic index $\gamma$, nested surfaces, and fluctuating profiles leads to an improved version of the standard MHD equilibrium principle found in popular codes such as VMEC and DESC.
The fluctuations act to regularize non-smooth solutions, leading to a predictable nondimensional fine length scale $\lambda$ (Eq.~\eqref{uniform_soln_second_order} and Fig.~\ref{fig:J-sheet}). 
Assuming $G$ is a diffeomorphism, we have shown that the statistical equilibrium model is elliptic when fluctuations appear in the profiles only (Eq.~\eqref{eq:statistical-ellipticity}). The question of ellipticity for more general fluctuation models remains open.

The statistical equilibrium model therefore improves upon the difficulties of the MHD equilibrium model while simultaneously including more physical effects. 
Unlike most attempts to rectify issues with the MHD equilibrium model, the statistical equilibrium model neither increases the complexity of computing solutions nor breaks the assumption of an average flux surface structure.
The only significant new requirement to pose the problem is the parameter $\lambda$, which is physically determined by the amplitude and length scale of the fluctuations according to Eq.\,\eqref{lambda_def}. In the context of equilibrium reconstruction from experimental diagnostics, $\lambda$ could be treated as a fitting parameter. In the context of stellarator design, $\lambda$ could be determined self-consistenty by iterating between a statistical equilibrium solver, like the one presented in Section \ref{sec:nonlinear_computation}, and a gyrokinetic turbulence code such as GENE \cite{jenkoElectronTemperatureGradient2000,gorlerGlobalVersionGyrokinetic2011,hoflerMilestonePredictingCore2025}.  

The linearized boundary layer analysis of the statistical equilibrium model in Section \ref{sec:asymptotics} shows that even though solutions become singular as $\lambda\rightarrow 0$, as expected from the MHD equilibrium model, the potential energy remains finite. In particular the $\lambda$-dependent part of the potential energy satisfies
\begin{equation}
    \frac{1}{2}\lambda^2 \int v_0'^2 \left(\left(\partial_{r}R\right)^2 + R^2 (n^2 + m^2)\right)\df^3\bm{x} \xrightarrow{\lambda \to 0} \frac{\lambda}{4\pi}\frac{\abs{v_0'}(R_+-R_{-})^2}{\sqrt{m^2 + n^2}}\abs{m\psi_P'(r_s)+n \psi_T'(r_s)}4\pi^2,
\end{equation}
where the integral over the singular layer may be estimated by observing that $\partial_{r}R$ in the singular layer is a Lorentz distribution, i.e., a mollified Dirac delta distribution, so $(\partial_{r}R)^2\sim \lambda^{-2}$ as $\lambda \to 0$. Despite the appearance of a Dirac delta current sheet in the small-$\lambda$ limit of the linear problem, we stress that the regularized current sheet is pressure-driven \cite{huang_structure_2023}. Delta-function current sheets, which arise from the kernel of the $\bm{\mathsf{B}}\cdot \nabla$ operator, are absent from the statistical equilibrium. This suggests that there may be a class of finite-energy weak solutions of the traditional MHD equilibrium model given as limits of solutions of the statistical equilibrium model, and that these solutions exclude delta-function current sheets.

As mentioned in Section \ref{sec:derivation}, with the minimal fluctuation model specified in Eq.\,\eqref{simple_fluctuation_model}, the general statistical equilibrium equations \eqref{statistical_equilibrium_divergence_form} simplify to Eqs.\,\eqref{statistical_equilibrium_v}-\eqref{statistical_equilibrium_zeta}. These simplified equations can also be written in the suggestive form
\begin{gather*}
    \mu_0^{-1}(\nabla\times\bm{\mathsf{B}})\times\bm{\mathsf{B}} + \mu_0^{-1}\frac{\delta \Psi_0^2}{\ell_0^6}(\nabla\times \bm{e}_T)\times\bm{e}_T + \mu_0^{-1}\frac{\delta \Psi_0^2}{\ell_0^6}(\nabla\times \bm{e}_P)\times\bm{e}_P = \nabla \mathsf{p},\\
     \bm{e}_T = \nabla v\times \nabla\theta,\quad \bm{e}_P = -\nabla v\times \nabla\zeta.
\end{gather*}
In light of Eqs.\,\eqref{statistical_equilibrium_theta}-\eqref{statistical_equilibrium_zeta}, the vector fields $\bm{e}_T,\bm{e}_P$ satisfy
\begin{align*}
    \nabla v\cdot \nabla\times\bm{e}_T & = -\nabla\cdot(\nabla v_\times \nabla v_\times \nabla\theta) = 0\\
    \nabla v\cdot \nabla\times\bm{e}_P & = \nabla\cdot(\nabla v_\times \nabla v_\times \nabla\zeta) = 0.
\end{align*}
It follows that both the mean current density $\mu_0^{-1}\nabla\times\bm{\mathsf{B}}$ and the mean magnetic field are tangent to flux surfaces. Equivalently, $(\mu_0^{-1}(\nabla\times\bm{\mathsf{B}})\times\bm{\mathsf{B}})\times\nabla \mathsf{p} =0$, which is a subset of the equations defining the MHD equilibrium model. Thus, relative to the MHD equilibrium model, statistical equilibrium only modifies radial force balance.  We remark that N. Sato previously studied $3D$ solutions of $(\mu_0^{-1}(\nabla\times\bm{\mathsf{B}})\times\bm{\mathsf{B}})\times\nabla \mathsf{p} =0$ in isolation in \cite{satoNestedInvariantTori2023}. Sato observed that this equation can be understood as a family of $2D$ elliptic equations parameterized by flux surface label. Since $(\mu_0^{-1}(\nabla\times\bm{\mathsf{B}})\times\bm{\mathsf{B}})\times\nabla \mathsf{p} =0$ is equivalent to the angular (i.e. $(\theta,\zeta)$) equations from either MHD or statistical equilibrium, Sato's family of $2D$ elliptic equations is equivalent to our Eqs.\,\eqref{break_degeneracy_zeta}-\eqref{break_degeneracy_theta}, whose surface-wise ellipticity played a crucial role in establishing ellipticity of the statistical equilibrium model.

In future work we will investigate the statistical equilibrium principle when $\bm{\mathsf B}_0$ has broken flux surfaces. 
To our knowledge, only the BETAS code \cite{betancourt_betas_1988} has attempted this problem in the MHD equilibrium context.
The statistical equilibrium model could greatly benefit the numerical solutions of this problem, including by smoothing singular currents and by relaxing the requirement that pressure be constant on magnetic field lines.
A numerical code based on this principle could be used to model non-integrable fields without the inclusion of resistive effects or time stepping, extending the fast prototyping and equilibrium reconstruction capabilities usually only afforded to nested solvers.

It would be interesting to investigate downstream implications of our modified equilibrium model, with an eye toward resolving apparent contradictions between theory, computation, and experiments. For instance, what explains the computation of precise quasisymmetric VMEC fields by Landreman-Paul \cite{landreman_magnetic_2022} decades after Garren-Boozer \cite{garren_existence_1991} argued that precise quasisymmetry can only be achieved on a single flux surface? It was shown that quasisymmetric MHD equilibria necessarily have zero singular Pfirsch-Schluter currents \cite{rodriguez_islands_2021}, but the nonlinear modifications and singular currents due to imperfect quasisymmetry require further analysis. The statistical equilibrium model, with its inherently smooth solutions, provides a logically-consistent starting point for revisions to near-axis expansion theory, as well as other aspects of advanced confinement concepts such as the coincidence of quasisymmetry and omnigeneity for analytic fields. The fact that the current remains parallel to surfaces implies the existence of Boozer coordinates for statistical equilibria \cite{rodriguez_generalized_2021}, allowing the for the direct use of the standard formulation of quasisymmetry. 

%

There is a thematic relationship between the statistical equilibrium model and the so-called Lagrangian-averaged MHD model due to Holm \cite{holmLagrangianAveragesAveraged2002}. Both models involve an ensemble average of a variational principle for ideal MHD. However, while Lagrangian-averaged MHD assumes flux freezing at the small scales that are ultimately averaged-out, statistical equilibrium assumes that small-scale physics causes flux-breaking. This conceptual difference reflects a difference in modeling  paradigms; Lagrangian-averaged MHD is designed as a turbulence closure for ideal MHD, while statistical equilibrium is intended as a closure for kinetic degrees of freedom, as is appropriate for stellarator modeling. This distinction has important mathematical consequences: Lagrangian-averaged MHD increases the order of the highest derivative in the equilibrium problem, while statistical equilibrium does not. Also in contrast the Lagrangian-averaged MHD model, in this work we have only implemented our statistical hypotheses for the equilibrium problem. It would be straightforward to develop the dynamical analogue of the theory in future work.

The methodology herein is mechanically similar to recent work on information geometric regularization (IGR) for Euler's equation \cite{cao_information_2026,barham_hamiltonian_2025}.
The benefit of IGR over other regularizations is that it is inviscid, allowing for smooth solutions of Euler's equation that do not dissipate energy.
Like the statistical equilibrium model, IGR is performed by modifying a variational problem defined on the space of diffeomorphisms.
It is interesting to consider whether a connection between the two could be constructed, allowing for an information geometric interpretation of the statistical equilibrium model.

The regularizing effects of the statistical equilibrium also opens the door to improved computational methodology. 
For instance, one could incorporate adaptive mesh refinement, such as has already been done in axisymmetry \cite{sanchez-vizuet_adaptive_2020,peng_adaptive_2020}, to resolve the $\lambda$-scale near rational surfaces.
Because the statistical equilibrium principle is elliptic, the Hessian of the variational principle is better behaved, suggesting that new preconditioning strategies could be developed for iterative solvers (e.g.,~through Eqs.~\eqref{break_degeneracy_theta} and \eqref{break_degeneracy_zeta}).
Even in the case of a MHD equilibrium solver, an ``artificial'' $\lambda$ could potentially be chosen as a function of the grid spacing, akin to artificial viscosity for resolving shocks in computational fluid dynamics.

\section{Acknowledgements}
The authors express their gratitude to E. Whal\'en, A. Bhattacharjee, D. Holm, H. Grayer, N. Cao, and J. Squire for helpful discussion during the preparation of this manuscript. This work was supported by US Department of Energy Contract DE-FG05-80ET-53088.


\bibliography{references.bib}

\appendix
\section{Degeneracy of the ideal MHD potential energy at rational surfaces\label{sec:math_motivation}}
The MHD potential energy functional $W(G)$ presents a major difficulty in developing a mathematical theory of 3D equilibria because it lacks \emph{coercivity}. Roughly speaking, coercive functionals take large values for large-norm arguments. 
Coercivity is an important ingredient for showing functionals like $W(G)$ have minimizers, from which well-posedness theory could be developed based on this basic existence result. 
However, the following heuristic argument shows that $W(G)$ cannot be coercive, and that therefore the existence problem for 3D equilibria is very delicate. 

Introduce a reference back-to-labels map $G=(v,\theta,\zeta)$ and a perturbation of that map $\overline{G} =(v,\overline{\theta},\overline{\zeta})$, where $\overline{\theta} = \theta + \iota(v)\chi$, $\overline{\zeta} = \zeta + \chi$, and $\chi$ is a smooth function on $Q$. We will use the reference map to help in constructing a sequence of $\chi$ functions with arbitrarily large derivatives such that $W(\overline{G})$ remains close to $W(G)$. The perturbation from $G$ to $\overline{G}$ does not change the energy stored in the magnetic field because it does not change the magnetic field itself. It does change the Jacobian according to $\overline{\mathcal{J}} = \mathcal{J} + \bm{B}\cdot\nabla\chi$, which implies a change in total internal energy for general $\chi$. Now consider the sequence of smooth $\chi$ functions indexed by the positive integer $N$: 
\begin{align*}
    \chi_N(v,\theta,\zeta) = \frac{1}{N^{\alpha-1}}w_0(N[v-v_r])\sin(N^\alpha\varphi),\quad \varphi=m\theta+n\zeta.
\end{align*}
Here $v_r$ denotes a resonant value of toroidal flux, $n+\iota(v_r)m = 0$, $\alpha >2$ is a positive integer, and $w_0:\mathbb{R}\rightarrow\mathbb{R}$ is a smooth non-negative bump function with support in the interval $(-1,1)$. It is important that $w_0$ and all of its derivatives vanish outside of the open interval $(-1,1)$. The $L^2$-norm of $\nabla\chi_N$ is given by
\begin{align*}
    ||\nabla\chi_N||^2 &= \int_{-1}^{1}\int_{0}^{2\pi}\int_{0}^{2\pi} \bigg( \frac{1}{N^{2(\alpha-2)}}w_0^\prime(X)^2\sin^2(N^\alpha\,\varphi)\,|\nabla v|^2\bigg)\frac{d\theta\,d\zeta\,dX}{N\mathcal{J}}\\
    & + \int_{-1}^{1}\int_{0}^{2\pi}\int_{0}^{2\pi} \bigg(\frac{1}{N^{\alpha-3}}w_0(X)w_0^\prime(X)\sin(N^\alpha\varphi)\cos(N^\alpha\varphi)\,\nabla\varphi\cdot\nabla v\bigg)\frac{d\theta\,d\zeta\,dX}{N\mathcal{J}}\\
    & + \int_{-1}^{1}\int_{0}^{2\pi}\int_{0}^{2\pi} \bigg(N^2\,w_0(X)^2\,\cos^2(N^\alpha\varphi)|\nabla\varphi|^2\bigg)\frac{d\theta\,d\zeta\,dX}{N\mathcal{J}}.
\end{align*}
As $N\rightarrow \infty$ the first two terms vanish, while the third term scales like $N$, implying $||\nabla\chi_N||^2\rightarrow \infty$ as $N\rightarrow \infty$. On the other hand, the potential energy of $\overline{G}$, assuming a barotropic equation of state for simplicity, is given by
\begin{align}
    W(\overline{G}) &=\frac{1}{2}\int|\bm{B}|^2\,d^3\bm{x}\nonumber\\
    &+ \int_{-1}^{1} \int_{0}^{2\pi}\int_{0}^{2\pi}\mathcal{U}(M^\prime\mathcal{J} + M^\prime\,\bm{B}\cdot\nabla\chi_N)(M^\prime\mathcal{J} + M^\prime\,\bm{B}\cdot\nabla\chi_N)\frac{d\theta\,d\zeta\,dX}{N\mathcal{J}}\nonumber\\
    & + \int_{0}^{v_r - 1/N}\int_{0}^{2\pi}\int_{0}^{2\pi}\mathcal{U}(\varrho)\,\varrho\,\frac{d\theta\,d\zeta\,dv}{\mathcal{J}}\nonumber\\
    &+ \int_{v_r+1/N}^{1}\int_{0}^{2\pi}\int_{0}^{2\pi}\mathcal{U}(\varrho)\,\varrho\,\frac{d\theta\,d\zeta\,dv}{\mathcal{J}},\label{non_coercive_W}
\end{align}
where the last three lines give the contributions to the total internal energy from the support of $w_0$, below the support of $w_0$, and above the support of $w_0$, respectively. We find
\begin{align*}
    \bm{B}\cdot\nabla\chi_N &= B^\zeta(n+\iota(v)m)N\,w_0(X)\,\cos(N^\alpha\varphi)\\
    & = B^\zeta m\iota^\prime(v_r)X\,w_0(X)\,\cos(N^\alpha\varphi) + N\,\mathcal{O}(X/N)^2.
\end{align*}
This implies that $\bm{B}\cdot\nabla\chi_N$ appearing on the second line of \eqref{non_coercive_W} is uniformly bounded in $N$ as $N\rightarrow\infty$. The second line in \eqref{non_coercive_W} therefore vanishes as $N\rightarrow \infty$, while the sum of the last two lines in \eqref{non_coercive_W} limits to the total internal energy of $G$. In other words
\begin{align*}
    \lim_{N\rightarrow\infty}W(\overline{G}) = W(G).
\end{align*}
We conclude that $W$ is not coercive.

It may seem simpler to establish non-coercivity of $W$ by choosing $\chi_N = \nu(Nv)$, where $C$ is some smooth single-variable function. Then $W(G) = W(\overline{G})$ exactly, independent of $N$, and $||\nabla\chi_N|| \rightarrow \infty$ as $N\rightarrow\infty$. While this is technically a valid argument, it is not interesting because shifting $\theta$ and $\zeta$ by flux functions corresponds to shifting the origin in the $(\theta,\zeta)$-plane on each flux surface. In other words, this form of perturbation corresponds to a redundancy in the representation of $\bm{B}$ and $p$ in terms of $G$, also known as a gauge symmetry. We may eliminate this redundancy by requiring that $(\theta,\zeta)$ vanishes at some conventional reference point on each flux surface. After imposing this restriction $\chi_N = \nu(Nv)$ no longer generates a valid perturbation of $G$ unless $C=0$. By contrast, the perturbation constructed above is \emph{not} a gauge transformation and still demonstrates non-coercivity.

It may also seem our argument for non-coercivity could be simplified dramatically by choosing $\chi$ to be $\delta$-localized on the rational surface. After all, the bump function $w_0(N[v-v_r])$ localizes strongly around the resonant surface as $N\rightarrow\infty$. We re-emphasize however that we are interested in smooth solutions of the equilibrium model. The behavior of $W(G)$ under non-smooth perturbations does not impinge directly on the smooth existence question.

Our chosen sequence $\chi_N$ represents a ``dead spot" for the functional $W$. Two features of the MHD potential energy functional conspire to create this dead spot. (a) Any perturbation of the angle variables $\theta,\zeta$ that aligns with unperturbed field lines leaves the magnetic energy unchanged. (b) Near an unperturbed rational surface there are smooth perturbations along the field lines that are approximately flux functions and vary rapidly across field lines. If either property, (a) or (b), was eliminated the dead spot would disappear. Property (a) follows from the fact that $\theta$ and $\zeta$ enter the magnetic energy only in the combination $\nabla\theta - \iota\nabla\zeta$, thus allowing perturbations in the toroidal magnetic field to cancel perturbations in the poloidal magnetic field. 

\end{document}